\documentstyle[twoside,epsf,psfig]{article}
\catcode`\@=11
\long\def\@makefntext#1{
\protect\noindent \hbox to 3.2pt {\hskip-.9pt
$^{{\eightrm\@thefnmark}}$\hfil}#1\hfill}               

\def\@makefnmark{\hbox to 0pt{$^{\@thefnmark}$\hss}}    
\def\ps@myheadings{\let\@mkboth\@gobbletwo
\def\@oddhead{\hbox{}
\rightmark\hfil\eightrm\thepage}
\def\@oddfoot{}\def\@evenhead{\eightrm\thepage\hfil
\leftmark\hbox{}}\def\@evenfoot{}
\def\sectionmark##1{}\def\subsectionmark##1{}}

\def\qed{\hbox{${\vcenter{\vbox{                        
   \hrule height 0.4pt\hbox{\vrule width 0.4pt height 6pt
   \kern5pt\vrule width 0.4pt}\hrule height 0.4pt}}}$}}

\def\bsc{{\sc a\kern-6.4pt\sc a\kern-6.4pt\sc a}}       
\def\bflatex{\bf L\kern-.30em\raise.3ex\hbox{\bsc}\kern-.14em 
T\kern-.1667em\lower.7ex\hbox{E}\kern-.125em X} 

\input{ijmpb1.sty}
\begin{document}

\runninghead{Raman Scattering in Cuprate Superconductors}{Raman Scattering in 
Cuprate Superconductors}

\normalsize\textlineskip
\thispagestyle{empty}
\setcounter{page}{1}

\copyrightheading{}                     {Vol. 0, No. 0 (1996) 000---000}

\vspace*{0.88truein}

\fpage{1}
\centerline{\bf RAMAN SCATTERING IN CUPRATE SUPERCONDUCTORS}
\vspace*{0.37truein}
\centerline{\footnotesize THOMAS PETER DEVEREAUX}
\vspace*{0.015truein}
\centerline{\footnotesize\it Department of Physics, George Washington 
University}
\baselineskip=10pt
\centerline{\footnotesize\it Washington, D.C. 20052, U.S.A.}
\vspace*{10pt}
\centerline{\normalsize and}
\vspace*{10pt}
\centerline{\footnotesize ARNO PAUL KAMPF}
\vspace*{0.015truein}
\centerline{\footnotesize\it Institut f\"ur Theoretische Physik, Universit\"at
zu K\"oln}
\baselineskip=10pt
\centerline{\footnotesize\it Z\"ulpicher Str. 77, 50937 K\"oln, Germany}
\vspace*{10pt}

\vspace*{0.21truein}
\abstracts{A theory for electronic Raman scattering in the cuprate
superconductors is presented with a specific emphasis
on the {\it polarization} dependence of the spectra which can infer the
symmetry of the energy gap. Signatures of the
effects of disorder on the low frequency and low temperature behavior of
the Raman spectra for different symmetry channels provide detailed
information about the magnitude and the phase of the energy gap. Properties
of the theory for finite T are discussed and compared to recent data
concerning the doping dependence of the Raman spectra in cuprate 
superconductors, and remaining questions are addressed.}{}{}

\vspace*{10pt}


\vspace*{1pt}\textlineskip      
\section{Introduction}    
\vspace*{-0.5pt}
\noindent
Interest in electronic Raman scattering has grown considerably since the
first theoretical analysis\cite{begth} in 1961
and the subsequent observation of 
the Raman effect in superconductors seventeen years ago.\cite{begexp} 
During the early- to mid- eighties
the amount of available data was limited to a few A-15 superconductors
while the theory was clarified in many aspects.\cite{theo} 
With the discovery of the
high temperature superconductors, this situation dramatically changed
as the amount of data grew on these systems. The theory of the Raman
effect in superconductors was completed by 1991 for both clean and disordered
s-wave superconductors,\cite{theoimp}
but this theory could not capture many features shown in experiments on
the cuprates. At present, experiments on the Raman effect exist for nearly all
``high'' temperature superconductors: materials which are
electron or hole-doped and materials with different number of CuO$_{2}$
planes.\cite{exp}

Recently attention has turned towards unconventional superconductivity 
candidates to describe the pairing in the cuprates.\cite{annett} 
Subsequently, due to the strong {\it symmetry} dependence of the
observed spectra, i.e., the
characteristic features of light scattering for different incident and
scattered polarization orientations, in all
high T$_{c}$ compounds, electronic Raman scattering in unconventional 
superconductors has grown to be of considerable interest in light of 
identifying the symmetry of
the energy gap in high temperature superconductors. Indeed, the amount of
attention lavished on this area has been remarkable and has provided a
large amount of detailed information towards understanding the mechanism of
pairing in these materials.  

The importance of Raman scattering can be
related to its ability to probe excitational dynamics on regions of the
Fermi surface rather than being restricted to measuring averages over the
Fermi surface. The symmetry dependence of the spectra
has augmented an understanding of the
magnitude and symmetry of the energy gap, and makes it as
effective a probe as photo-emission has proven to be 
to determine $\Delta({\bf k})$. Moreover Raman scattering is a bulk probe
of a material due to the long wavelength of the exciting laser light, does not
suffer appreciably from surface effects, and has extremely sharp energy
resolution.

Simple considerations of the transformation properties of the light scattering
amplitude were used in Ref. 7 to demonstrate how the light 
polarization orientations selectively probe excitational dynamics of regions 
of the Fermi surface or the Brillouin zone, and subsequent work has clarified 
this picture considerably.\cite{sub,tpdde,dvz} In the normal state, the light 
scattering cross section provides information concerning the scattering rate 
of electrons in certain ${\bf k}$-space regions. $B_{1g}$ scattering geometry 
[$(x-y)(x+y)$ orientation of incoming (scattered) light polarizations, 
respectively] probes excitations along the Brillouin zone axes 
while $B_{2g}$ [$(x)(y)$] probes the diagonals. 
Thereby ${\bf k}$-dependent scattering processes are measured simply by 
rotating the polarization orientations of the incoming and scattered photons. 
In the superconducting state, a ${\bf k}$ dependent energy gap can be 
mapped out. The direct coupling of the Raman vertex  
to an anisotropic energy gap $\Delta({\bf k})$ leads to a 
strong polarization dependence of the Raman spectra in the superconducting 
state. For an energy gap of $d_{x^{2}-y^{2}}$ symmetry, which is minimal 
along the zone
diagonals and maximal along the axes, the 
$B_{2g}$ orientation probes Cooper pairs
where the nodes of the gap are located and does not sample regions where the 
gap is maximal. The reverse is true for the
$B_{1g}$ orientation. Thus the 
$B_{2g}$ spectrum will reflect a smaller energy needed by the incoming light 
to break a Cooper pair and $B_{1g}$ will have a larger energy scale. Lastly, 
since $A_{1g}$ involves more of an average around the Brillouin Zone the 
energy scale can be less than that for $B_{1g}$. Moreover, screening effects 
for the fully symmetric charge distributions ($A_{1g}$) can drastically 
reorganize the spectrum for this channel as the long range Coulomb interaction 
is brought into play.

In most cases, the experimental Raman spectra for optimally
hole-doped cuprates can be well modeled using an energy gap of $d_{x^{2}-
y^{2}}$ symmetry. The relative peak positions, low frequency power-laws
and temperature
dependence of the nearly elastic contribution for $B_{1g}$ and $B_{2g}$
scattering channels can be naturally explained assuming a $d_{x^{2}-y^{2}}$
paired superconducting state. The telling signature here is the presence of an
$\omega^{3}$ frequency dependence of the $B_{1g}$ channel frequencies compared 
to a linear dependence in other channels. This uniquely identifies the nodal 
positions of the gap to lie along $45^{\circ}$ in the CuO$_{2}$ plane. 
However, it is to be noted that a small linear in frequency contribution is 
seen in the $B_{1g}$ channel in optimally doped YBa$_{2}$Cu$_{3}$O$_{7-\delta}$
(Y-123) which may be due to orthorhombic distortions\cite{strohm} which break 
the $B_{1g} - A_{1g}$ symmetry distinction and/or may be due to the large Fano 
distortion of the background due to the 340 cm$^{-1}$ out-of-phase oxygen 
vibration.\cite{dvz2} 
For the $A_{1g}$ channel (which can not be purely selected in one 
particular geometry) the strong peak of the observed spectrum can still be fit 
by an appropriate choice of the gap and scattering amplitude but the 
theoretical prediction was found to be extremely sensitive to the number of 
harmonics used to represent the ${\bf k}$-dependence and thus to small
changes in band structure and/or small dopings. Moreover screening 
necessarily leads in the calculations to a much smaller signal than seen in 
experiments. It may be that additional physics is needed to describe the 
$A_{1g}$ peak shape. However, the linear frequency behavior at small 
frequencies follows naturally due to a gap with nodes. As in most correlation 
functions for clean superconductors, only the absolute 
magnitude of the energy gap and not the phase is measured in Raman scattering.
The current agreement of the calculations for clean $d_{x^{2}-y^{2}}$
superconductors compared to the data on optimally doped 
Bi$_{2}$Sr$_{2}$CaCu$_{2}$O$_{8+\delta}$ (Bi-2212) are presented in 
Fig. 1.

\begin{figure}[htbp]
\vspace*{13pt}
\psfig{file=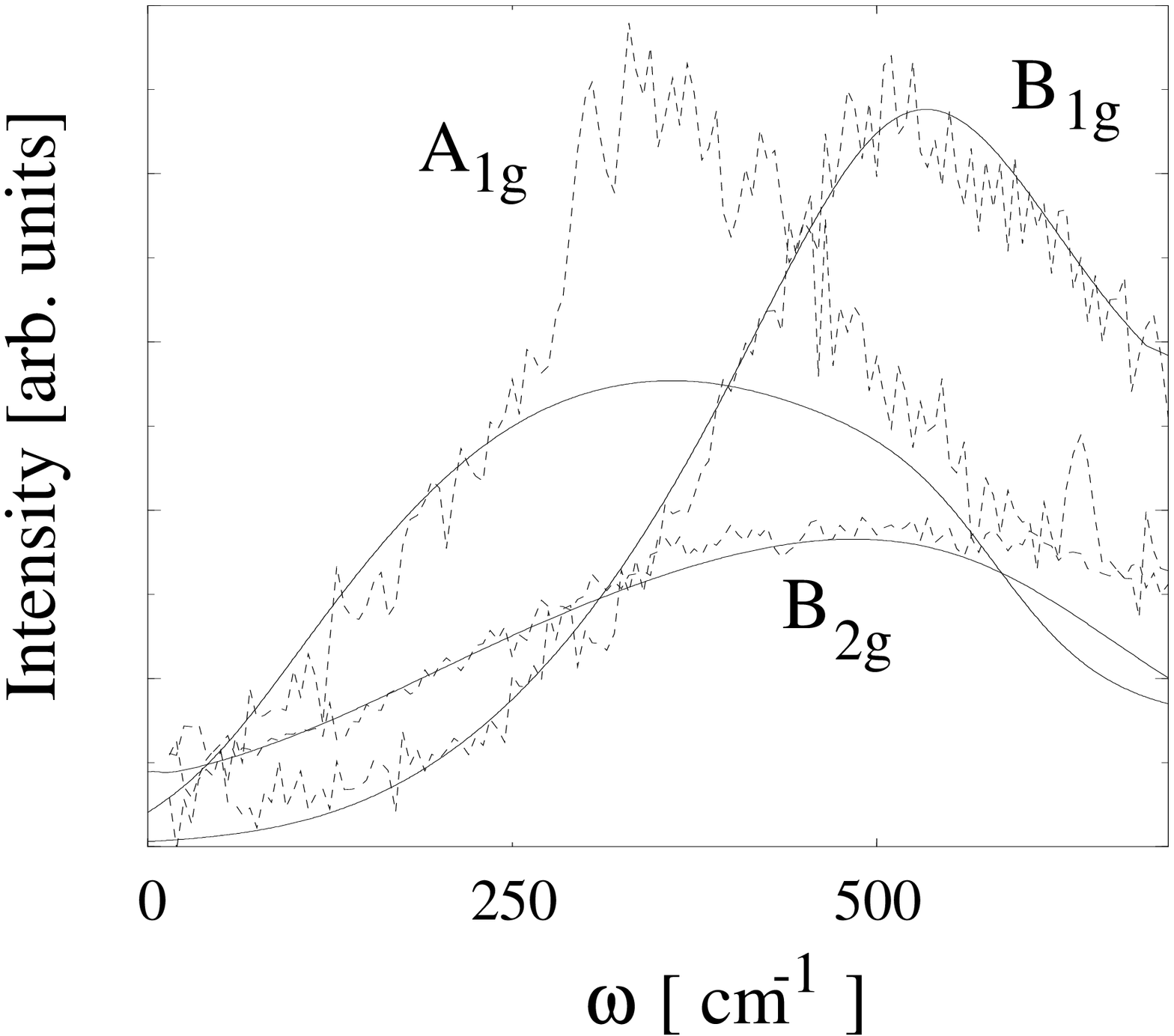,height=6.cm,width=8.cm,angle=0}
\fcaption{Comparison of the calculations\cite{tpdde} performed for Raman 
scattering in clean $d_{x^{2}-y^{2}}$ superconductors compared to the data 
taken on optimally doped Bi-2212 by Staufer et al. in Ref. 5.}
\label{fig. 1}
\end{figure}

However, the theory cannot link a description valid for all 
temperatures to the normal
state. For vanishing momentum transfers of the light, phase space for 
electronic Raman scattering can only be opened due to scattering via 
impurities or electron-electron or electron-phonon interactions. In the 
superconducting state this phase space restriction is lifted since breaking 
Cooper pairs requires no net momentum transfer.

Building upon the well known results of impurity effects in anisotropic 
$s$-wave superconductors,\cite{hoh} Borkowski and Hirschfeld\cite{bork} 
have also shown that while gap anisotropy is
unaffected by impurity scattering in $d-$wave superconductors,
systematic impurity doping in anisotropic $s-$wave superconductors leads to 
thermally activated behavior as the gap anisotropy 
is averaged out and can thus provide an indirect way of determining
if the gap has accidental nodal points or if the nodes are enforced by
symmetry. Since smearing effects can obscure
thermally activated behavior, an observation of a threshold in energy, 
such as the density of states (DOS), could also be useful. However, it is well 
known that DOS measurements can be problematic due to surface conditions and 
small coherence lengths.

Moreover, many open questions are encountered in cuprate superconductors
for the Raman spectra away from
optimal doping\cite{doping}. In overdoped materials a small 
region (if any) of cubic
behavior is seen in the $B_{1g}$ channel before crossing over to a linear
low frequency dependence. Strong scattering ($\sim \omega$) is seen down to 
the lowest frequencies measured in all overdoped materials for all other 
channels. In underdoped materials, difficulties are encountered to observe
features associated with superconductivity in the $B_{1g}$ channel while strong
scattering is still seen at low frequencies for all channels and is 
linear in frequency as well.\cite{irwin} 
Additionally the loss of overall scattering intensity of the $B_{1g}$ channel
relative to other channels has been taken as evidence for a 
pseudogap in the normal state.\cite{irwin} This may be
explained as the effect of losing part of the Fermi surface near $(\pi,0)$ as 
hole pockets near $(\pi/2,\pi/2)$ are developing. 
In addition, very distinct features and behaviors
are seen in resonant Raman scattering when the incoming laser light is 
varied, and behavior reminiscent of two-magnon scattering in an
antiferromagnetic insulator is seen to persist well into the underdoped
region of superconductivity.\cite{magnon1}
Therefore, while the existence of spectral
weight down to the lowest frequencies implies the existence of a minimum
energy gap at most as large as the energy resolution of the experiments, the
strong symmetry dependence of the spectra as a function of doping can contain
clues to the behavior of excitational dynamics and ultimately a mechanism of
pairing for the superconducting state.

While a complete theory now exists for Raman scattering in clean unconventional
superconductors, this paper is devoted to augmenting the theory to include
the effects of impurity scattering to exploit the fact that the response of 
a material to deliberate disorder can
yield more information concerning the anisotropy of the energy gap but
its phase as well. 
In this paper we will review the theory of Raman scattering in 
unconventional disordered superconductors with the goal of addressing
how the scattering cross section in the superconducting state can provide
information on the ground state symmetry as a function of doping. Open
questions will be addressed and comparison to the available data will be
made.

This paper is outlined as follows:
In Section 2 we shall review the essentials of the theory of Raman scattering
in metals and confine our attention to the case of only one band crossing
the Fermi level. The relation of the symmetry of the Raman vertex and its
coupling to an anisotropic energy gap is discussed.

In Section 3 we will present a theory of Raman scattering in disordered
unconventional superconductors. Isotropic $s-$wave impurity scattering
is included in a gauge invariant way for superconductors with arbitrary
gap symmetry at arbitrary temperatures.

Section 4 presents the channel dependent spectra obtained for
a $d_{x^{2}-y^{2}}$ and an anisotropic $s-$wave superconductor neglecting
vertex corrections and are compared and contrasted. The symmetry aspects 
of the calculations as a function of frequency and temperature
are discussed in detail. We also discuss the role of vertex corrections. 
In particular the existence of disorder generated collective modes are 
examined.

Lastly, Section 6 compares the results of the calculations to spectra
obtained on the cuprates at various dopings. The current agreement and lack
of agreement of the data with the theory are discussed, and open questions
are presented. A brief account of this work has appeared in Ref. 18.

\section{General formalism for metals}

Electronic Raman scattering measures effective charge fluctuations around
the Fermi surface, 
\begin{equation}
\tilde\rho=\sum_{{\bf k},\sigma} \gamma({\bf k,q=0})
c^{\dagger}_{{\bf k},\sigma}c_{{\bf k},\sigma} \label{one}
\end{equation}
with a scattering amplitude given by the Raman vertex,
\begin{eqnarray}
&&\gamma({\bf k,q=0};\omega_{I},\omega_{S})={\bf e^{I} \cdot e^{S}}+ 
{1\over{m}}\sum_{\nu}\\
&&\times\left[
{\langle n,{\bf k}\mid {\bf e^{S} \cdot p}\mid\nu ,{\bf k}\rangle  
\langle \nu ,{\bf k}\mid {\bf e^{I} \cdot p}\mid n,{\bf k}\rangle   
\over{\epsilon({\bf k})-\epsilon_{\nu}({\bf k})+\omega_{I}}}+
{\langle n,{\bf k}\mid {\bf e^{I} \cdot p}\mid\nu ,{\bf k}\rangle  
\langle \nu ,{\bf k}\mid {\bf e^{S} \cdot p}\mid n,{\bf k}\rangle   
\over{\epsilon({\bf k})-\epsilon_{\nu}({\bf k})-\omega_{S}}}\right],
\nonumber \label{two}
\end{eqnarray}
where ${\bf e^{I}, e^{S}} (\omega_{I},\omega_{S})$ are the incident and 
scattered photon polarization
vectors (energies), ${\bf p}=-i\hbar{\bf\nabla}$, and $\epsilon({\bf k})$ and
$\epsilon_{\nu}({\bf k})$ are the Bloch conduction and intermediate state
energies, respectively.\cite{theo} While in the limit of vanishing light 
frequencies
and for a single band near the Fermi level the Raman scattering amplitude 
can be related to the curvature of the conduction band, in general this
relation does not hold and is of questionable use for the cuprates.
A thorough discussion of the Raman vertex is given in the Appendix of 
Ref. 10. An alternative approach is based on the experimental 
observation that the spectra near optimal doping in the normal state depends
only mildly on the incoming laser light. Then the scattering amplitude can
be taken as roughly independent of frequency and symmetry considerations can
be given to its ${\bf k}-$dependence. This of course misses resonant Raman
scattering which will be more relevant for smaller dopings nearer to the
antiferromagnetic phase. However, it is a much more complex and unresolved 
problem of how to bridge non-resonant to resonant scattering.

Therefore we elect to describe the Raman vertex $\gamma$ in terms of an
expansion of Fermi surface (FS) or Brillouin zone (BZ) harmonics.
The vertices depend on momentum throughout the BZ as:
\begin{eqnarray}
&&B_{1g}:~~~~\gamma({\bf k}) \sim \cos(k_{x}a)-\cos(k_{y}a) + \cdots\\
&&B_{2g}:~~~~\gamma({\bf k}) \sim \sin(k_{x}a)\sin(k_{y}a) +\cdots \nonumber \\
&&A_{1g}:~~~~\gamma({\bf k}) \sim \rm{constant} + \cos(k_{x}a)+\cos(k_{y}a) +
\cdots, \nonumber
\label{three}
\end{eqnarray}
for a 2-D lattice with lattice constant $a$. Here $\cdots$ represent higher 
order terms in the BZ expansion. This allows us to correctly
classify the anisotropy and transformation properties of the scattering
amplitude but leaves open the question of absolute intensities due to
the unknown momentum independent prefactors of the expansion. The prefactors
can be taken as parameters to fit overall intensities. Apart from $A_{1g}$
(see Ref. 10) the prefactors have only a small effect on the frequency 
lineshape of the spectra.

The channel-dependent Raman cross section
is related to the channel-dependent Raman susceptibility 
$\chi_{\gamma,\gamma}$ via the fluctuation-dissipation theorem,
\begin{eqnarray}
\frac{\partial^2\sigma}{\partial\omega\partial\Omega}&=&
\frac{\omega_{\rm S}}{\omega_{\rm I}}\ r_0^2\ S_{\gamma,\gamma}({\bf q},\omega),
\nonumber \\
S_{\gamma,\gamma}({\bf q},\omega)&=&-\frac{1}{\pi}
\left[1+n(\omega)\right]{\rm Im}\ \chi_{\gamma,\gamma}({\bf q},\omega),
\label{four}
\end{eqnarray}
with 
\begin{equation}
\chi_{\gamma,\gamma}(i\omega)=\int_{0}^{1/T} d\tau e^{-i\omega\tau}
\langle T_{\tau}[\tilde\rho(\tau)\tilde\rho]\rangle, \label{five}
\end{equation}  
with $T_{\tau}$ the time-ordering operator and the imaginary part is
obtained by analytic continuation, $i\omega \rightarrow \omega +i0$. 
Here $r_0=e^2/mc^2$ is the Thompson radius and
we have set $\hbar=k_{B}=1$.

In the absence of impurity scattering, the Raman response function can be
written in terms of the ${\bf k}-$dependent Tsuneto function
\begin{equation}
\lambda({\bf k}, i\omega)=
{\Delta({\bf k})^{2}\over{E({\bf k})^2}}
\tanh\biggl[{E({\bf k})\over{2 T}}\biggr]
\left[{1\over{2E({\bf k})+i\omega}}+
{1\over{2E({\bf k})-i\omega}}\right].\label{six}
\end{equation}
as
\begin{equation}
\chi_{\gamma,\gamma}(i\omega)=
\sum_{{\bf k}}\gamma^{2}({\bf k})\lambda({\bf k},i\omega), \label{seven}
\end{equation}
with the excitation energy $E^{2}({\bf k})=\xi^{2}({\bf k})+
\Delta^{2}({\bf k})$, conduction band $\xi({\bf k})=\epsilon({\bf k})-\mu$, 
$\mu$ the chemical potential, and energy gap $\Delta({\bf k})$. 
We have neglected Coulomb screening
(important for fully symmetric charge fluctuations $A_{1g}$) as well as
pair interactions responsible for maintaining gauge invariance and
collective modes. For details of calculations performed with Eq. 
(\ref{seven}), the reader is referred to Refs. \cite{tpd,sub,tpdde,dvz}.

From Eqs. (\ref{six}-\ref{seven}), we see that the Raman vertex
couples directly to the energy gap under the momentum summation. Since
the $B_{1g}$ channel assigns maximum weight along the BZ
axes ($0,\pm 1$) and ($\pm 1,0$), while $B_{2g}$ weights the diagonals 
($\pm 1,\pm 1$),
the orientation of the incident and scattering light polarizations thus
choose effective charge fluctuations on the corresponding regions of the
FS. Therefore, if for instance a $d_{x^{2}-y^{2}}$ energy
gap is used, $\Delta({\bf k})=\Delta_{0}[\cos(k_{x}a)-\cos(k_{y}a)]/2$, 
the $B_{1g} (B_{2g})$
channel samples regions of the gap maximum (minimum), respectively. The 
combination of the two symmetries thus gives information about the nodal
behavior as well as the maximum value of the energy gap $\Delta_{0}$. 

\section{Disordered unconventional superconductors}

\subsection{{\mbox{\boldmath$T$}}-matrix approach}

We now consider the effect of scattering by spinless, noninteracting,
isotropic impurities on the Raman susceptibility of unconventional
superconductors. We use the self consistent
$T-$matrix approach to incorporate repeated 
scattering events from a single impurity site and dress both the Green's 
functions as well as the vertex. 

The two parameters in this
theory are the cotangent of the scattering phase shift,
$c=\cot (\delta)$, and the impurity concentration $n_{i}$
described through the scattering rate $\Gamma=(N/V)n_{i}/\pi N_{F}$,
where $N/V$ is the electron density and $N_{F}$ is the density of states per
spin at the Fermi level.\cite{hew} These
enter into the $\hat T$-matrix and the self energy in particle-hole (Nambu)
space:
\begin{equation}
\hat \Sigma({\bf k},i\omega)= \Gamma \hat T({\bf k,k},i\omega),
\label{eight}
\end{equation}
The $\hat T$ matrix (in terms of the self energy) is depicted in Fig. 2 and 
satisfies a Bethe-Salpeter equation,
\begin{equation}
\hat T({\bf k,p},i\omega)=\hat V({\bf k,p})+\sum_{\bf k^{\prime}}
\hat V({\bf k,k^{\prime}})\hat G({\bf k^{\prime}},i\omega)
\hat T({\bf k^{\prime},p},i\omega),
\label{nine}
\end{equation}
with $\hat G$ the Green's function in Nambu space which contains the
$\hat T$ matrix via Dyson's equation. The matrix 
$\hat V({\bf k,p})$ is
the impurity scattering potential for a single electron, taken to be 
independent of the electron's spin. Moreover, the potential is taken to
be independent of momentum as well to represent $s-$wave scattering only.
Therefore the self energy is ${\bf k}-$ independent as well.

\begin{figure}[htbp]
\vspace*{13pt}
\psfig{file=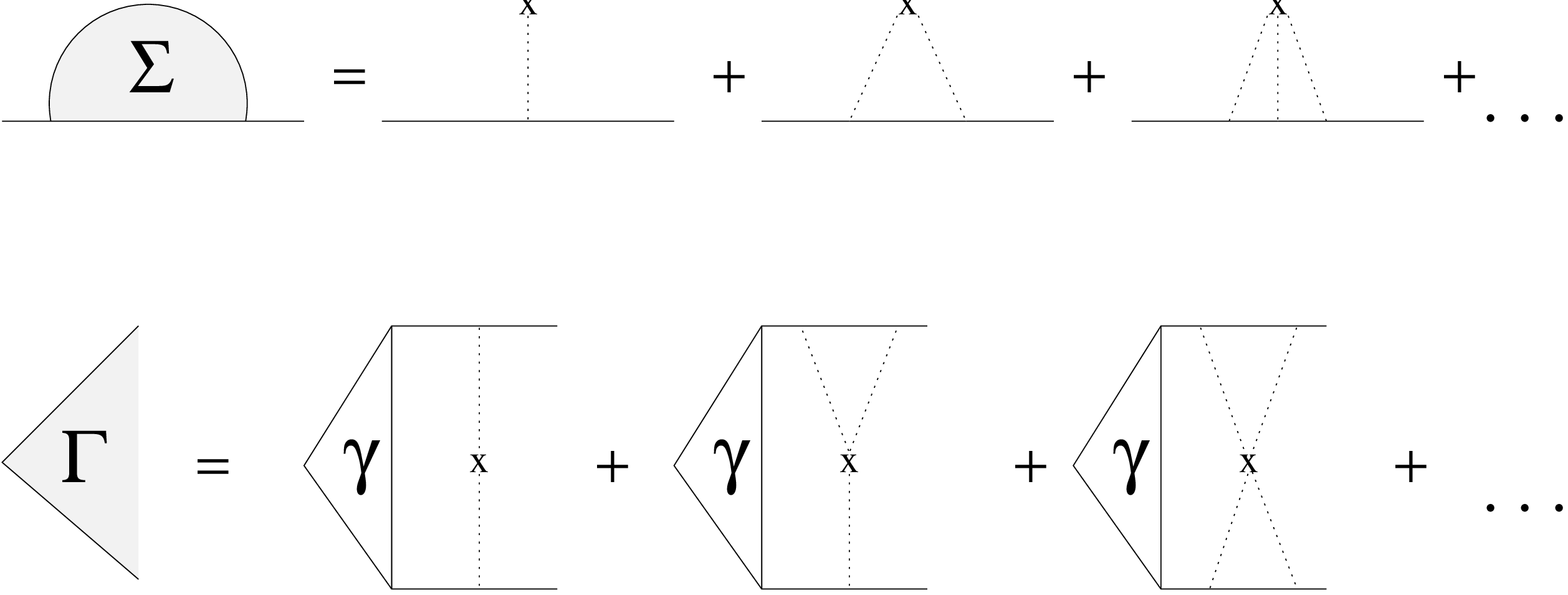,height=4.cm,width=8.cm,angle=270}
\vspace*{0.4truein}             
\fcaption{Diagrams for the T-matrix self energy $\Sigma$ and the
renormalized vertex $\Gamma$ as defined in the text.}
\label{fig. 2}
\end{figure}

Expanding the self energy in quarternions, 
$\hat \Sigma = \sum_{i=0}^{3} \hat\tau_{i} \Sigma_{i}$, where $\hat\tau_{0}$
is the $2\times 2$ unit matrix and $\hat\tau_{i} (i=1,2,3)$ are the Pauli
matrices, the one-particle Green's function can be written as
\begin{equation}
\hat G({\bf k},i\omega)={i\tilde\omega\hat\tau_{0}+
\tilde\xi({\bf k})\hat\tau_{3}
+\tilde\Delta({\bf k})\hat\tau_{1}
\over{(i\tilde\omega)^{2}-\tilde\xi^{2}({\bf k})-\tilde
\Delta^{2}({\bf k})}}.
\label{ten}
\end{equation}
The tilde indicates the renormalized frequency, gap, and band 
energy via
\begin{equation} 
i\tilde\omega=i\omega-\Sigma_{0}(i\tilde\omega),~~
\tilde\Delta({\bf k})=\Delta({\bf k})+\Sigma_{1}(i\tilde\omega),~~ 
\tilde\xi({\bf k})=\xi({\bf k})-\Sigma_{3}(i\tilde\omega).
\label{eleven}
\end{equation}
The matrix self energy is given in terms of the integrated
Green's function $g_{i}(i\omega)={1\over{2\pi N_{F}}}\sum_{{\bf k}}Tr 
\{\hat\tau_{i}\hat G({\bf k}, i\omega)\}$ as
\begin{equation}
\hat\Sigma(i\omega)=\Gamma{g_{0}(i\omega)\hat\tau_{0}-
g_{1}(i\omega)\hat\tau_{1}-c\hat\tau_{3}\over{
c^{2}-g_{0}^{2}(i\omega)+g_{1}^{2}(i\omega)}}=\Sigma_{0}\hat\tau_{0}+
\Sigma_{1}\hat\tau_{1}+\Sigma_{3}\hat\tau_{3}.
\label{twelve}
\end{equation}
Here $Tr$ denotes taking the trace. The off-diagonal self energy $\Sigma_{1}$
is zero only for odd-parity states or states which possess reflection
symmetry such as $d_{x^{2}-y^{2}}$ or $d_{xy}$. Here we have assumed 
particle-hole symmetry and further set $g_{3}(i\omega)=0$. This allows us
to cast the self energy in the simplified form given in Eq. (\ref{twelve}).
As discussed in Ref. 19, this approximation is valid for either weak
(Born, $c>>0$) or strong (unitary, $c=0$) scattering provided that 
impurity vertex corrections play only a minor role. This is shown to be the
case in Section 4.3, and therefore we use Eq. (\ref{twelve}) for the self
energy. For a more detailed discussion of this point, the reader is 
referred to Ref. 19.

The expression for the Raman response in the limit of vanishing
momentum transfer is given by
\begin{equation}
\chi_{\gamma,\gamma}({\bf q}\rightarrow 0,i\Omega)= 
-T\sum_{i\omega} \sum_{\bf k}Tr \{\gamma({\bf k}) \hat \tau_{3} 
\hat G({\bf k}, i\omega^{+}) \hat\tau_{3}
\hat\Gamma({\bf k},i\tilde\omega^{+},i\tilde\omega^{-})
\hat G({\bf k},i\omega^{-})\}, 
\label{thirteen}
\end{equation}
where $\omega^{\pm}=\omega\pm\Omega/2$, and
$\gamma\hat\tau_{3}, \hat\Gamma$ are the bare 
and impurity renormalized Raman vertices, respectively.
The diagrammatic representation for the renormalized vertex is shown in 
Fig. 2 and can be expressed as the integral equation
\begin{equation}
\hat\Gamma({\bf k},i\tilde\omega^{+},i\tilde\omega^{-})
=\gamma({\bf k})\hat\tau_{3}+\sum_{\bf p}
\hat T^{+}\hat G({\bf p},i\tilde\omega^{+})
\hat\Gamma({\bf p},i\tilde\omega^{+},i\tilde\omega^{-})
\hat G({\bf p},i\tilde\omega^{-})\hat T^{-},
\label{fourteen}
\end{equation}
with $\hat T^{\pm}=\hat T(i\tilde\omega^{\pm})$. 

Eqs. (8-14) form a closed set of equations for the Raman response 
of unconventional superconductors in the self consistent $\hat T$-matrix 
approximation. We remark that while the impurities are included in a gauge 
invariant way, the neglect of the pairing interaction in the renormalized 
vertex does not satisfy gauge invariance and therefore all information 
regarding the existence of pairing interaction induced collective modes and 
sum rules is lost. While a gauge invariant
treatment is possible for disordered $s$-wave superconductors\cite{theoimp}
a similar treatment has not yet been performed for unconventional 
superconductors.
We note that for clean superconductors, the gauge invariant Raman response 
has been calculated in Ref. 9
for $d-$ wave superconductors. There it was found that the collective modes 
which do exist in different Raman channels are damped and lead only to small 
shifts in the relative peak positions of the Raman spectra in each channel. 
The low frequency behavior in particular is unaffected. However the 
Goldstone mode which appears as a consequence of the spontaneously broken
$U(1)$ gauge symmetry, provides for both longitudinal
and transverse screening of the $A_{1g}$ response in a non-trivial way, as 
discussed in length in Ref. 9. Therefore we are not in a position 
to discuss the behavior of the $A_{1g}$ Raman response in the presence of 
impurities for unconventional superconductors since this would require a gauge 
invariant treatment.

\subsection{Solution of the integral equations}

To solve Equation (\ref{fourteen}), it is convenient to first remove the
${\bf k}-$ dependent term and define
$\hat\Gamma({\bf k},i\omega,i\omega^{\prime})=\hat\tau_{3}\gamma({\bf k})+
\hat\delta(i\omega,i\omega^{\prime})$, and then expand once again
in quarternions, $\hat\delta=\sum_{i=0}^{3}\hat\tau_{i}\delta_{i}$. Eq.
(\ref{fourteen}) then turns into a $4\times 4$ matrix integral equation with
no non-zero elements for general $\Delta({\bf k})$. To simplify a solution,
we now restrict attention to odd-parity states or states which possess 
reflection symmetry such as $d_{x^{2}-y^{2}}$ or $d_{xy}$. Therefore
the energy gap satisfies $\sum_{\bf k}\Delta({\bf k})=0$ and the off-diagonal
term of the self energy $\Sigma_{1}$ can be set to zero. 
The resulting matrix equation
then simplifies to
\begin{eqnarray}
&&\delta_{0}=\sum_{\bf k}[
I_{00}L^{00}({\bf k})\delta_{0}+I_{03}(L^{33}({\bf k})\gamma({\bf k})+
L^{33}({\bf k})\delta_{3})]\nonumber \\
&&\delta_{1}=\sum_{\bf k}[
I_{11}L^{11}({\bf k})\delta_{1}-I_{12}(L^{23}({\bf k})\gamma({\bf k})+
L^{22}({\bf k})\delta_{2})]\nonumber \\
&&\delta_{2}=\sum_{\bf k}[
I_{11}(L^{23}({\bf k})\gamma({\bf k})+L^{22}({\bf k})\delta_{2})+
I_{12}L^{11}({\bf k})\delta_{1}]\nonumber \\
&&\delta_{3}=\sum_{\bf k}[
I_{00}L^{33}({\bf k})(\gamma({\bf k})+\delta_{3})+
I_{03}L^{00}({\bf k})\delta_{0}],
\label{fifteen}
\end{eqnarray}
where we have simplified notation and defined the following:
\begin{eqnarray}
&&I_{00}=T_{0}^{+}T_{0}^{-}+T_{3}^{+}T_{3}^{-}, ~~~
I_{11}=T_{0}^{+}T_{0}^{-}-T_{3}^{+}T_{3}^{-}, \nonumber \\
&&I_{03}=T_{0}^{+}T_{3}^{-}+T_{3}^{+}T_{0}^{-}, ~~~
I_{12}=-i(T_{0}^{+}T_{3}^{-}-T_{3}^{+}T_{0}^{-}), ~~~
\label{sixteen}
\end{eqnarray}
and
\begin{eqnarray}
&&L^{00}({\bf k})=
{i\tilde\omega^{+}i\tilde\omega^{-}+\Delta^{2}({\bf k})+\tilde\xi^{+}({\bf k})\tilde\xi^{-}({\bf k})\over{N^{+}({\bf k})N^{-}({\bf k})}}, \nonumber \\
&&L^{11}({\bf k})=
{i\tilde\omega^{+}i\tilde\omega^{-}+\Delta^{2}({\bf k})-\tilde\xi^{+}({\bf k})\tilde\xi^{-}({\bf k})\over{N^{+}({\bf k})N^{-}({\bf k})}},\nonumber \\
&&L^{22}({\bf k})=
{i\tilde\omega^{+}i\tilde\omega^{-}-\Delta^{2}({\bf k})-\tilde\xi^{+}({\bf k})\tilde\xi^{-}({\bf k})\over{N^{+}({\bf k})N^{-}({\bf k})}},\nonumber \\
&&L^{33}({\bf k})=
{i\tilde\omega^{+}i\tilde\omega^{-}-\Delta^{2}({\bf k})+\tilde\xi^{+}({\bf k})\tilde\xi^{-}({\bf k})\over{N^{+}({\bf k})N^{-}({\bf k})}},\nonumber \\
&&L^{23}({\bf k})={i\Delta({\bf k})[i\tilde\omega^{+}-i\tilde\omega^{-}]
\over{N^{+}({\bf k})N^{-}({\bf k})}}, 
\label{seventeen}
\end{eqnarray}
with
$N^{\pm}({\bf k})=(i\tilde\omega^{\pm})^{2}-(\tilde\xi^{\pm}({\bf k}))^{2}-
\Delta^{2}({\bf k})$. 

The matrix equation can then be solved for general vertex $\gamma({\bf k})$.
Denoting $\langle A({\bf k})\rangle =\sum_{\bf k} A({\bf k})$, we obtain
the solution to the vertex equation as
\begin{eqnarray}
&&\delta_{0}=I_{03}{\langle\gamma({\bf k})L^{33}({\bf k})\rangle+
\delta_{3}\langle L^{33}({\bf k})\rangle\over{1-I_{00}
\langle L^{00}({\bf k})\rangle}},\nonumber \\
&&\delta_{1}=-I_{12}{\langle\gamma({\bf k})L^{23}({\bf k})\rangle+
\delta_{2}\langle L^{22}({\bf k})\rangle\over{1-I_{11}
\langle L^{11}({\bf k})\rangle}},\nonumber \\
&&\delta_{2}=\langle\gamma({\bf k})L^{23}({\bf k})\rangle
{I_{11}-{I_{12}^{2}\langle L^{11}({\bf k})\rangle\over{1-I_{11}\langle 
L^{11}({\bf k})\rangle}}\over{1-I_{11}\langle L^{22}({\bf k})\rangle
+I_{12}^{2}{\langle L^{11}({\bf k})\rangle \langle L^{22}({\bf k})
\rangle\over{1-I_{11}\langle L^{11}({\bf k})\rangle}}}}, \nonumber \\
&&\delta_{3}=\langle\gamma({\bf k})L^{33}({\bf k})\rangle
{I_{00}+{I_{03}^{2}\langle L^{00}({\bf k})\rangle \over{1-
I_{00}\langle L^{00}({\bf k})\rangle}}\over{1-I_{00}\langle L^{33}({\bf k})
\rangle - I_{03}^{2}{\langle L^{00}({\bf k})\rangle \langle L^{33}({\bf k})
\rangle\over{1-I_{00}\langle L^{00}({\bf k})\rangle}}}}
\label{eighteen}
\end{eqnarray}
The conserving approximation for the vertex equation (\ref{fourteen}) enforces
that the impurities have been treated in a gauge invariant way for the
the renormalized vertex. However, the full theory is not gauge invariant
since the pairing interactions responsible for superconductivity have not been
included in the vertex equation.

\subsection{Raman susceptibility}

We are now in a position to obtain our final result for the Raman 
susceptibility. Substituting Eq. (\ref{eighteen}) into Eq.
(\ref{thirteen}) and performing the trace, the Raman susceptibility
follows as
\begin{eqnarray}
&&\chi({\bf q}=0,i\Omega)=2T\sum_{i\omega}\Biggl\{
\langle \gamma^{2}({\bf k})L^{33}({\bf k})\rangle \nonumber \\
&&+\langle \gamma({\bf k})L^{33}({\bf k})\rangle^{2}
{I_{00}+I_{03}^{2}{\langle L^{00}({\bf k})\rangle\over{1-I_{00}\langle L^{00}({\bf k})
\rangle}}\over{1-I_{00}\langle L^{33}({\bf k})\rangle-I_{03}^{2}
{\langle L^{00}({\bf k})\rangle \langle L^{33}({\bf k})\rangle\over{1-I_{00}
\langle L^{00}({\bf k})\rangle}}}}\nonumber \\
&&-\langle \gamma({\bf k})L^{23}({\bf k})\rangle^{2}
{I_{11}+I_{12}^{2}{\langle L^{11}({\bf k})\rangle\over{1-I_{11}\langle L^{11}({\bf k})
\rangle}}\over{1-I_{11}\langle L^{22}({\bf k})\rangle +I_{12}^{2}
{\langle L^{11}({\bf k})\rangle \langle L^{22}({\bf k})\rangle\over{1-I_{11}
\langle L^{11}({\bf k})\rangle}}}}\Biggr\}.
\label{nineteen}
\end{eqnarray}
The first term is the response in the absence of vertex corrections while
the next two terms are due to vertex corrections.
To obtain the final result for the Raman cross section, one must
analytically continue by letting $i\Omega \rightarrow \Omega +i0$ and
take the imaginary part of Eq. (\ref{nineteen}) to be put into
Eq. (\ref{four}). 

We now consider the role of symmetry in the vertex corrected response.
We focus on $A_{1g}$ scattering first to show its connection to
gauge invariance.
For isotropic density fluctuations, $\gamma({\bf k})= \rm{constant}$, and thus
by symmetry the last term in Eq. (\ref{nineteen}) is zero. Part of the
second term is canceled by the first term and we are still left with a
finite response at $q=0$ for isotropic density fluctuations. This is due to
having a theory which breaks gauge invariance and violates the
$f-$sum rule and thus particle number conservation. The density response
will be made to vanish when the pair interaction is included in the
vertex renormalization and when long-range screening by the Coulomb forces
are taken into account. This only affects the results for fully symmetric
scattering ($A_{1g}$).
We note that for general scattering in the $A_{1g}$ channel, 
$\gamma({\bf k})$ need not be a constant. However, by symmetry we see that
the third term in Eq. (\ref{nineteen}) vanishes for $A_{1g}$ 
scattering in $d-$wave superconductors.

Next we consider $B_{1g}$ and $B_{2g}$ scattering channels. Again by
symmetry we see that the second term vanishes for both channels. However,
the third term contributes for channels with the same symmetry as the
energy gap $\Delta({\bf k})$. For a gap with $d_{x^{2}-y^{2}}$ pairing,
the third term is zero for $B_{2g}$ scattering but contributes for $B_{1g}$.
The opposite is true for $d_{xy}$ pairing. Thereby we see that the vertex
corrections are very symmetry dependent and can by themselves lead to 
channel-dependent line-shapes.

In the following section we will discuss the results of the theory in
the absence of vertex corrections and at zero temperature
to bring out the differences between
the results from clean unconventional superconductors. In particular we
will investigate how impurity effects can help determine the phase of
the energy gap. We defer discussions of the role of vertex corrections
to the following section.

\section{Scattering in the absence of vertex corrections}

\subsection{{\mbox{\boldmath$T$}}=0 results}

In this section we will calculate the effects of impurities on the Raman
spectra for unconventional superconductors by evaluating Eqs. 
(\ref{seventeen}-\ref{nineteen}).
To simplify the calculations, in what follows we will work with an approximately
cylindrical FS and restrict all momentum averages to the
FS. The vertices then depend on azimuthal angle $\phi$ around
the FS as
\begin{eqnarray}
&&B_{1g}:~~~~\gamma(\phi) = \sqrt{2}\cos(2\phi)\\
&&B_{2g}:~~~~\gamma(\phi) = \sqrt{2}\sin(2\phi),\nonumber
\label{twenty}
\end{eqnarray}
where we also neglect higher order terms. 
We also neglect band structure details and for simplicity take an
infinite band as well. We thus approximate all ${\bf k}$ sums as an energy
integral times an angular integral around the FS:
$$\sum_{\bf k} \rightarrow N_{F} \int d\xi d\Omega_{\bf k}.$$
The reader is referred to Ref. 9 for a discussion of consequences of the these
approximations. For instance we are no longer able to discuss the role of the
van Hove singularity. Recent calculations show that the van Hove singularity 
due to the flat bands located near $(\pi,0)$ shows up as a peak in 
the $B_{1g}$ Raman spectrum
at an energy $\sim \sqrt{\Delta^{2}+\epsilon_{vH}^{2}}$, where $\epsilon_{vH}$
is the distance in energy of the flat band from the Fermi level near the
zone axis (see Ref. 10 and Branch and Carbotte in Ref. 8). 
However no such van Hove peak, which should be sensitive to band
structure and should not be tied to the energy gap, is seen in experiments. 
Since the van Hove is at higher energies
for most systems it is likely that quasiparticle inelastic scattering is
sufficiently strong to damp this feature if it exists, or it may be lying at 
further distances away from the Fermi level and thus have inappreciable 
residue. In any case we neglect this since impurities are sufficient 
to suppress the van Hove feature in any case. 

In this section we consider specifically
a $d_{x^{2}-y^{2}}$ paired superconductor, with energy gap
$\Delta_{d}(\phi)=\Delta_{0}\cos(2\phi)$ and compare with the
results for a hypothetical anisotropic $s-$wave superconductor with
$\Delta_{s}(\phi)=\Delta_{0}\mid\cos(2\phi)\mid$. In the absence of impurity
scattering, the channel-dependent Raman spectra (as well as the results for
other correlation functions) would be indistinguishable. We will consider
only the $T=0$ response and defer consideration of finite temperatures as
well as vertex corrections to the next subsection and following
section, respectively. Since $\sum_{\bf k} \Delta({\bf k})$ is not zero for
the anisotropic $s$ case, we reinsert the $\Sigma_{1}$ term into $L^{33}$
($\Delta$ is replaced by $\tilde\Delta$ in Eq. (\ref{seventeen}))
when evaluating integrals for the vertex uncorrected calculations. This is
the only modification needed.

\begin{figure}[htbp]
\vspace*{13pt}
\psfig{file=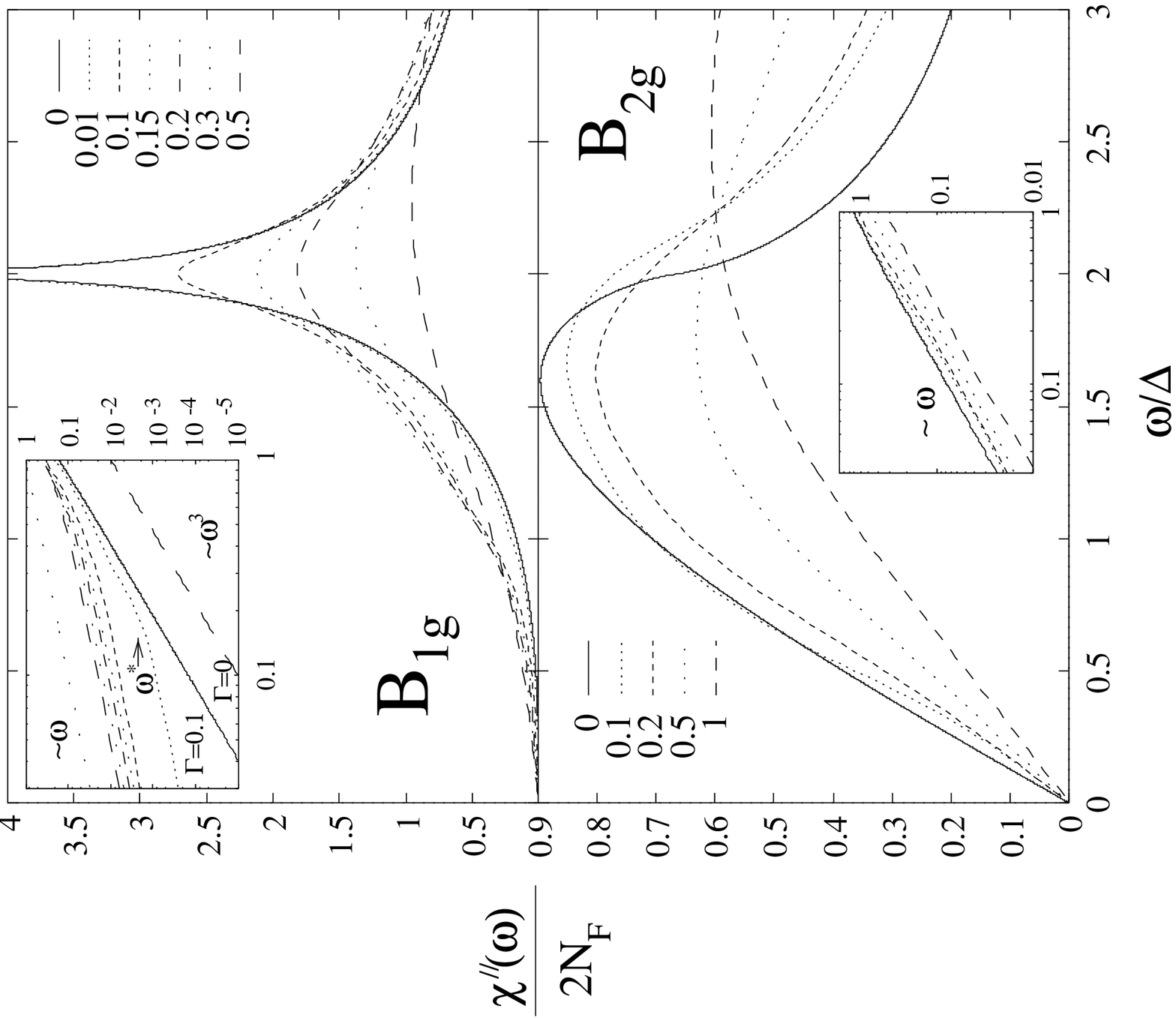,height=10.cm,width=8.cm,angle=270}
\fcaption{Raman spectra for $B_{1g}$ and $B_{2g}$ channels for 
$d_{x^{2}-y^{2}}$ pairing for $T=0$. Values of $\Gamma/\Delta$ are shown for
resonant scattering. The inset shows the low frequency behavior on a log-log
scale. The $B_{1g}$ spectra show a cross over from $\omega$ to $\omega^{3}$
at roughly $\omega^{*}\sim \sqrt{\Gamma\Delta}$. The bounding lines 
in the $B_{1g}$ insets are guides 
for $\omega$ (upper) and $\omega^{3}$ (lower) behavior.}
\label{fig. 3}
\end{figure}

The results of the
calculations for unitary scattering $c=0$ are shown in Figs. 3 and 4
for the $B_{1g}$ and $B_{2g}$ channels for both superconductors. One
immediately notices that the impurities smear out the logarithmic divergence
of the peak at $2\Delta_{0}$ for the $B_{1g}$ channel for both
superconductors. Moreover, as the scattering rate $\Gamma/\Delta_{0}$
increases past roughly 0.5, the differences between the peak of the
spectra in both channels become less pronounced as the scattering becomes
the dominant energy scale in the problem. 
But as a consequence of the gap renormalization for anisotropic $s$-wave 
superconductors, $\Sigma_{1}$ is non zero and
the gap becomes averaged out by the disorder and a 
threshold develops at $\omega_{g}=2\Delta_{min}$. For small impurity
scattering, this leads to a reduction
of the relative peak positions for the $B_{1g}$ and $B_{2g}$ channels
compared to $d_{x^{2}-y^{2}}$, as shown in Figs. 3 and 4. 
As the disorder is increased,
the peak positions will coalesce and the spectra recover a channel
independent, isotropic $s-$wave form, \cite{theoimp} 
$$\chi^{\prime\prime}_{s-wave}(\omega) \sim \Theta(\omega-2\Delta_{ave})
{\Delta_{ave}\over{\Gamma}}
\ \rm{for} \ \ \omega \sim 2\Delta_{ave}$$
with $\Delta_{ave}=2\Delta_{0}/\pi$. 

\begin{figure}[htbp]
\vspace*{13pt}
\psfig{file=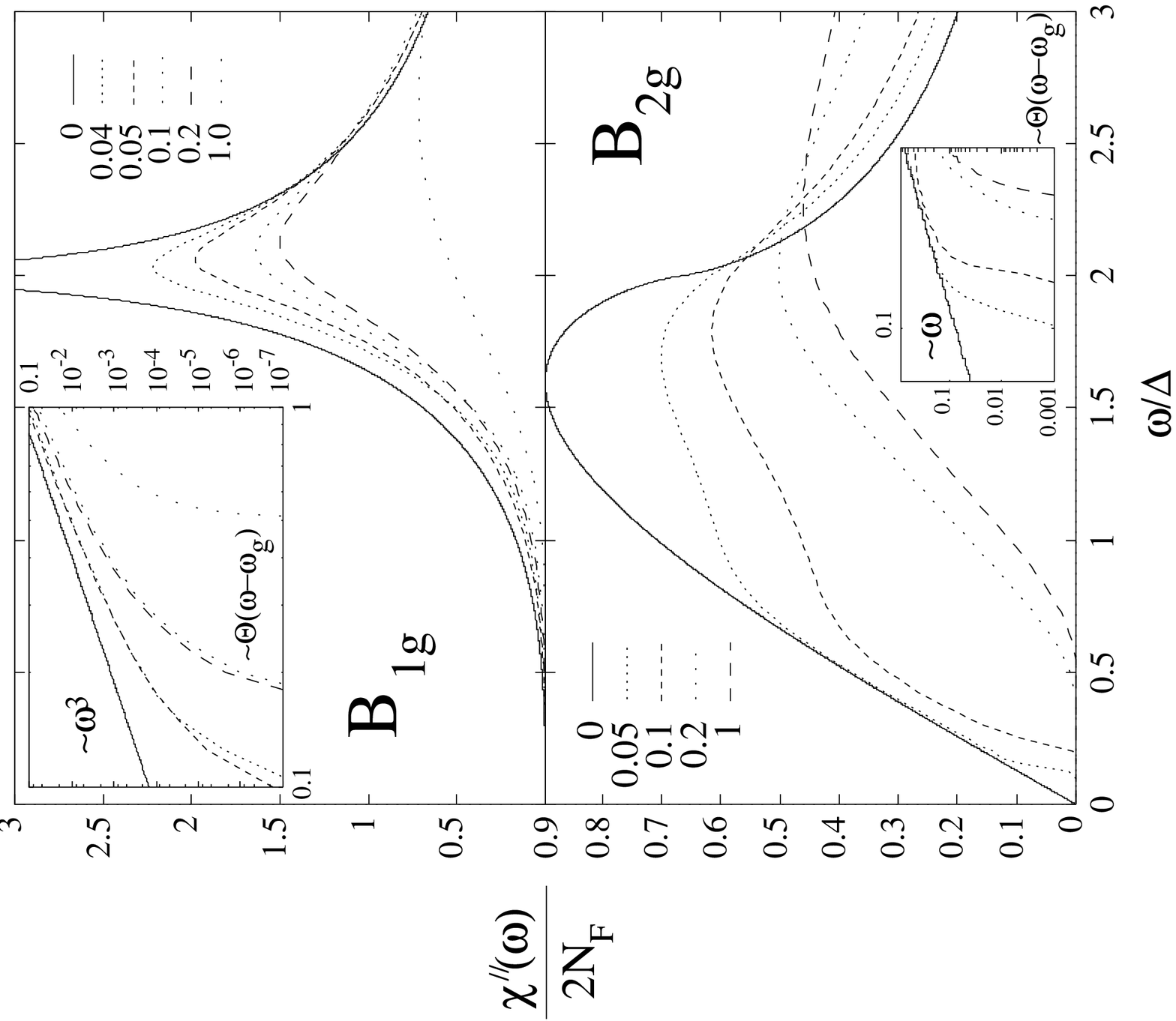,height=10.cm,width=8.cm,angle=270}
\fcaption{Raman spectra for $B_{1g}$ and $B_{2g}$ channels for a 
superconductor with a $\mid d_{x^{2}-y^{2}}\mid $ gap symmetry. Inset shows
the low frequency behavior which is dominated by a threshold for both 
channels.}
\label{fig. 4}
\end{figure}

However, the main difference for $d_{x^{2}-y^{2}}$ and anisotropic
$s$-pairing lies in the low frequency behavior, which is very channel
dependent, as shown in the insets of Figs. 3 and 4. 
For clean materials, both superconductors show the characteristic
$\omega (\omega^{3})$ behavior for the $B_{2g} (B_{1g})$ channels, 
respectively.

For $d_{x^{2}-y^{2}}$ pairing, while impurities do not change the 
linear in frequency behavior for the $B_{2g}$ channel, 
below a characteristic frequency $\omega^{*} \sim \sqrt{\Gamma \Delta}$ the 
behavior crosses over from $\omega^{3}$ to $\omega$ for the $B_{1g}$ 
channel. This is due to a nonzero density of states at the Fermi level, 
which allows for normal-state-like behavior to be recovered. \cite{gorkov}
As in the case of the penetration depth,\cite{pjh} the scale
$\omega^{*}$ grows with increasing impurity concentrations, and will be
shown in the next section to be strongly temperature dependent.
However, {\it the exponent is symmetry dependent} 
(remains 1 for $B_{2g}$ and $A_{1g}$ channels,
while {\it decreases} from 3 to 1 for the $B_{1g}$ channel).
This is in marked contrast to the impurity dependence of the
spectra for anisotropic $s$ gap as seen in Fig. 4. 
The low frequency behavior is dominated by the threshold in this case
for {\it all channels}. Moreover, the impurity dependence of the 
$B_{1g}$ channel is opposite to what one would expect if the gap was 
anisotropic $s$-wave. For anisotropic $s$, one sees a transfer of spectral
weight out to higher frequencies due to the development of the threshold 
while for
$d_{x^{2}-y^{2}}$ the transfer of spectral weight is towards lower frequencies.
Thus impurity scattering can provide a clear
qualitative way of distinguishing energy gaps of different symmetry.

It can be shown that Born scattering ($c>>0$) leads to the same effects
except for low frequencies. The spectra reorganize at higher frequencies
($\omega \sim 2\Delta_{0}$) in the same way as for resonant scattering. Also,
as has been pointed out\cite{bork} there is little difference between Born
and unitary scattering for anisotropic $s-$pairing even at low frequencies.
However, the low frequency behavior is changed for $d-$pairing. In the Born
limit, the low frequency behavior in the $B_{1g}$ channel
crosses over from $\omega$ to $\omega^{3}$
at a much smaller frequency $\omega^{*}$. Therefore the low frequency
behavior of the spectra is similar to the behavior obtained
in the absence of impurities except for only very small frequencies in the
$B_{1g}$ channel. A substantial linear behavior for $B_{1g}$ can be obtained
in the Born limit for large scattering, $\Gamma/\Delta_{0} \sim 1+c^{2}$,
but such a large scattering almost completely smears out the peak feature at
$2\Delta_{0}$.

For small scattering, the threshold generated in anisotropic $s-$ wave
superconductors can be obscured
via inelastic scattering or experimental resolution. 
However, the ``effective'' exponents for {\it all} channels would grow as the 
threshold develops for increased impurity scattering.
Since low temperatures and low frequencies can be achieved
at high resolution ($\sim 8$ cm$^{-1}$), the smearing can be controlled to
allow for a systematic check of the impurity dependence
to determine whether the gap has accidental or intrinsic zeroes.

\subsection{Finite temperatures}

We now consider the effect of finite temperatures on the calculations.
The same physics that is manifest in the symmetry dependent low frequency
behavior gives rise to a similar characteristic temperature dependence
which can also be used to ascertain the energy gap symmetry.

We first note that for temperatures above T$_{c}$, the Raman spectrum
for channel $L$ is given by a simple Lorentzian:
\begin{equation}
\chi^{\prime\prime}_{L}(\Omega,T>T_{c})=2N_{F} 
\gamma_{L}^{2}{\Omega\tau\over{(\Omega\tau)^{2}+1}},
\label{twentyone}
\end{equation}
where $\tau=1/2\Gamma$ is the impurity scattering lifetime.\cite{normimp} 
In the presence of vertex corrections $\tau$ becomes channel dependent.
This spectrum peaks at frequencies $\Omega=1/\tau$, rises linearly with
frequency and falls off at large frequencies as $1/\Omega$. More
complicated forms for $\tau$ can be obtained for ${\bf k}-$dependent 
scattering
and when electron-electron interactions are taken into account.\cite{eeraman}

Our results recover this form as $T$ approaches $T_{c}$. Therefore, any
cubic rise of the $B_{1g}$ channel must shrink as temperatures increase
and the peak in the spectra must approach the normal state value as well.
Both factors lead to a temperature dependence of the crossover frequency
$\omega^{*}$ estimated in the previous subsection.

\begin{figure}[htbp]
\vspace*{13pt}
\psfig{file=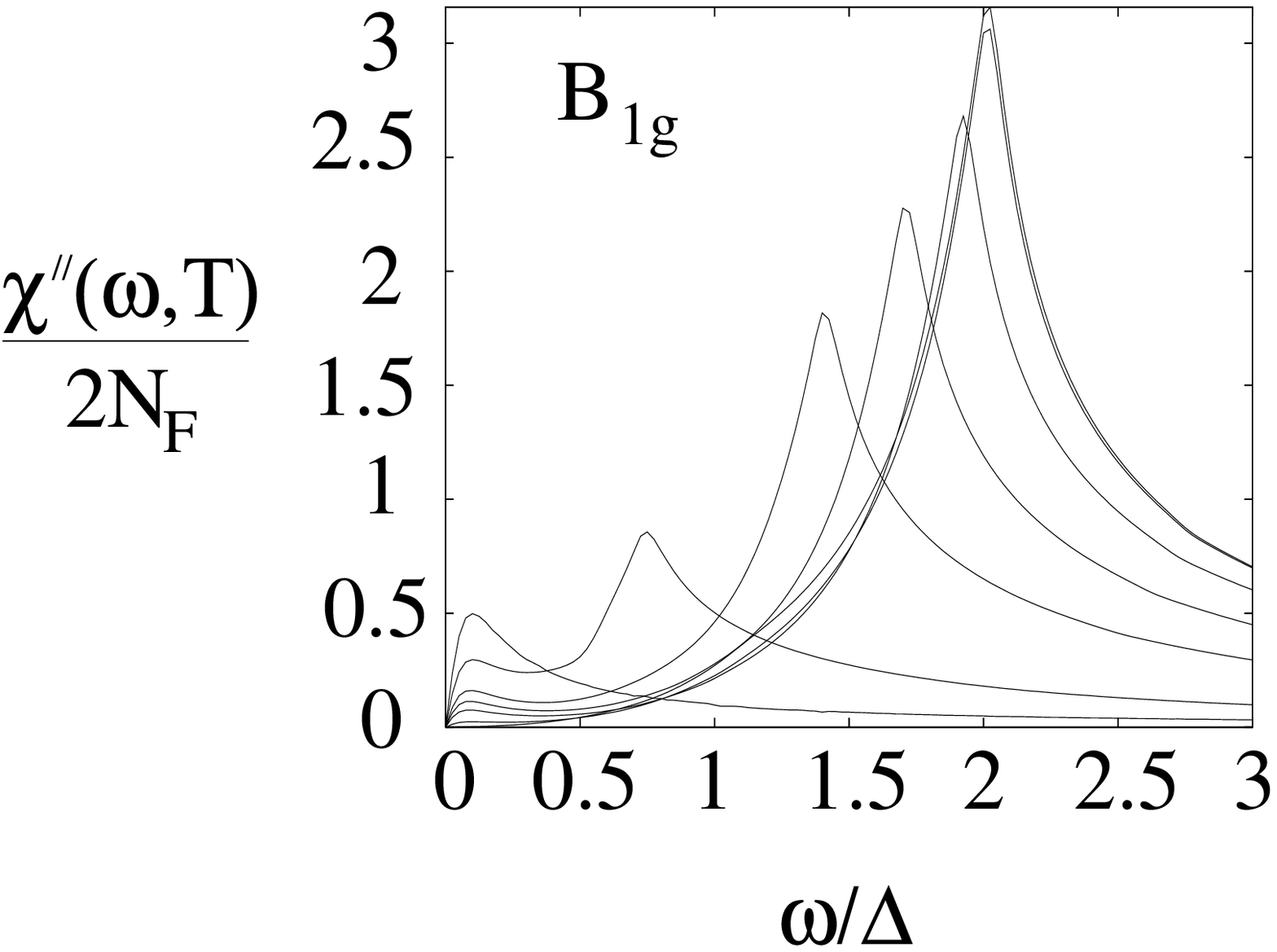,height=6.cm,width=8.cm,angle=270}
\fcaption{$B_{1g}$ Raman spectrum evaluated for
$\Gamma/\Delta=0.05$ and for unitary scatterers at successive temperatures
$T/T_{c}=0.1,0.5,0.8,0.9,0.99$ and $1$, respectively. Here a weak coupling 
temperature dependence of the energy gap $\Delta/T_{c}=2.14$\cite{tpdde} has 
been assumed for simplicity.}
\label{fig. 5}
\end{figure}

The temperature dependence of the results for $d_{x^{2}-y^{2}}$
superconductors are shown in Figs. 5 and 6. Here we have
assumed a weak coupling temperature dependent energy gap for simplicity,
whose weak coupling value is $2\Delta_{0}/T_{c}=4.28$. 
From the Figure it can be seen how the normal
state Lorentzian line-shape rises out of the superconducting response
as the temperature increases and the energy gap decreases. An additional
peak appears at $\Omega=1/\tau$ in both channels
with little weight at low temperatures
and increasing in magnitude as the temperature increases. The low frequency
behavior is thus dramatically changed so that
the spectrum can easily loose any trace of cubic behavior if the impurity
scattering and/or the temperature is large enough. Therefore only at very low
temperatures can an elastic scattering rate be inferred from the data. 

\begin{figure}[htbp]
\vspace*{13pt}
\psfig{file=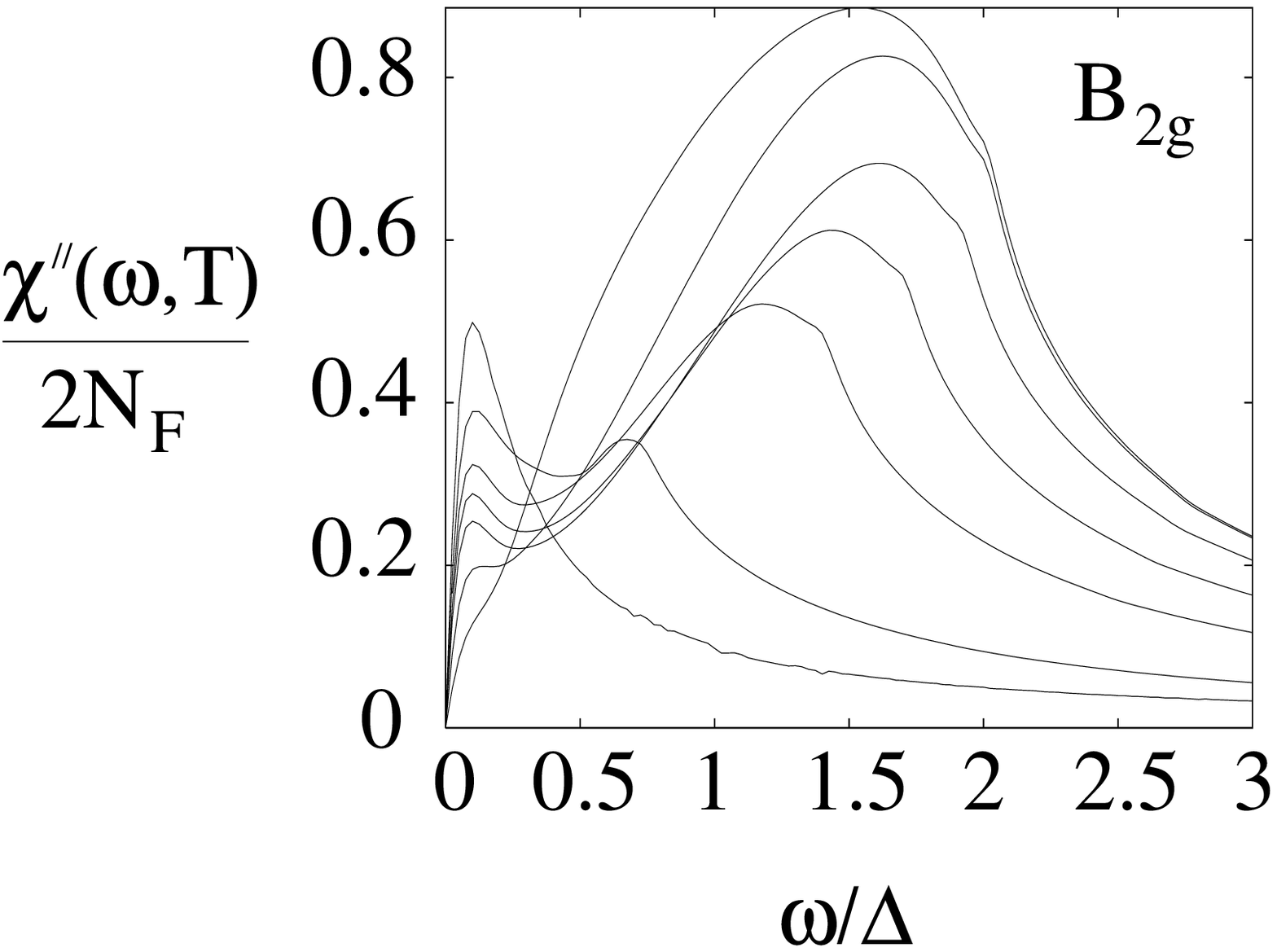,height=6.cm,width=8.cm,angle=0}
\fcaption{$B_{2g}$ temperature dependent spectrum using the same parameters
as in Fig. 5.}
\label{fig. 6}
\end{figure}

Additional information can be obtained by observing the low frequency
behavior as a function of temperature. The ratio of the superconducting
response to the normal state response in the static limit $\Omega
\rightarrow 0$ shows characteristic temperature power-law behavior in the
case of clean $d-$wave superconductors as discussed in Refs. 7 and 9.
There it was shown that for clean materials the ratio is given by a
weighted average over the FS of a Fermi function $f$ as
\begin{eqnarray}
{\chi^{\prime\prime}_{s.s.}(\Omega \rightarrow 0,T)\over{
\chi^{\prime\prime}_{n.s.}(\Omega \rightarrow 0,T)}}&=& 
{2\langle f(\mid\Delta({\bf k})\mid)\mid\gamma({\bf k})\mid^{2}\rangle\over{
\langle\mid\gamma({\bf k})\mid^{2}\rangle}},~~~~\rm{clean}\nonumber \\
&&\sim T^{3}, ~~~~~~~ T\rightarrow 0, ~~~~~B_{1g},\nonumber \\
&&\sim T, ~~~~~~~ T\rightarrow 0, ~~~~~B_{2g},
\label{twentytwo}
\end{eqnarray}
where the last two lines gives the same exponents for the low temperature
behavior as the exponents for the frequency dependence. 
It can be seen that the spectra
develop a suppression of states at low frequencies faster for the $B_{1g}$
channel than $B_{2g}$, which has been borne out in experiments on clean
cuprate samples.\cite{exp} 

In the presence of disorder this simple expression does not hold. This is
due to the fact that the Green's functions do not have an undamped
simple pole any longer due to the pair breaking nature of impurities in
a $d-$wave superconductor. Moreover,
for clean systems above T$_{c}$ there is no Raman scattering so this limit
is in a sense artificial. The inclusion of impurity scattering verifies
the above form if the limit of vanishing scattering is taken as to make
the definition sensible. However, for finite scattering the static limit
ratio is not given by Eq. (\ref{twentyone}). For $d_{x^{2}-y^{2}}$
pairing, since for all temperatures below T$_{c}$ the $B_{2g}$ and 
$A_{1g}$ channels have a linear dependence on $\Omega$ for small frequencies 
while even $B_{1g}$ has a small linear dependence as well, this linear
dependence mimics the normal state behavior and thus gives a constant term
even at $T=0$. Therefore all curves will be shifted upwards at low 
temperatures.

\begin{figure}[htbp]
\vspace*{13pt}
\psfig{file=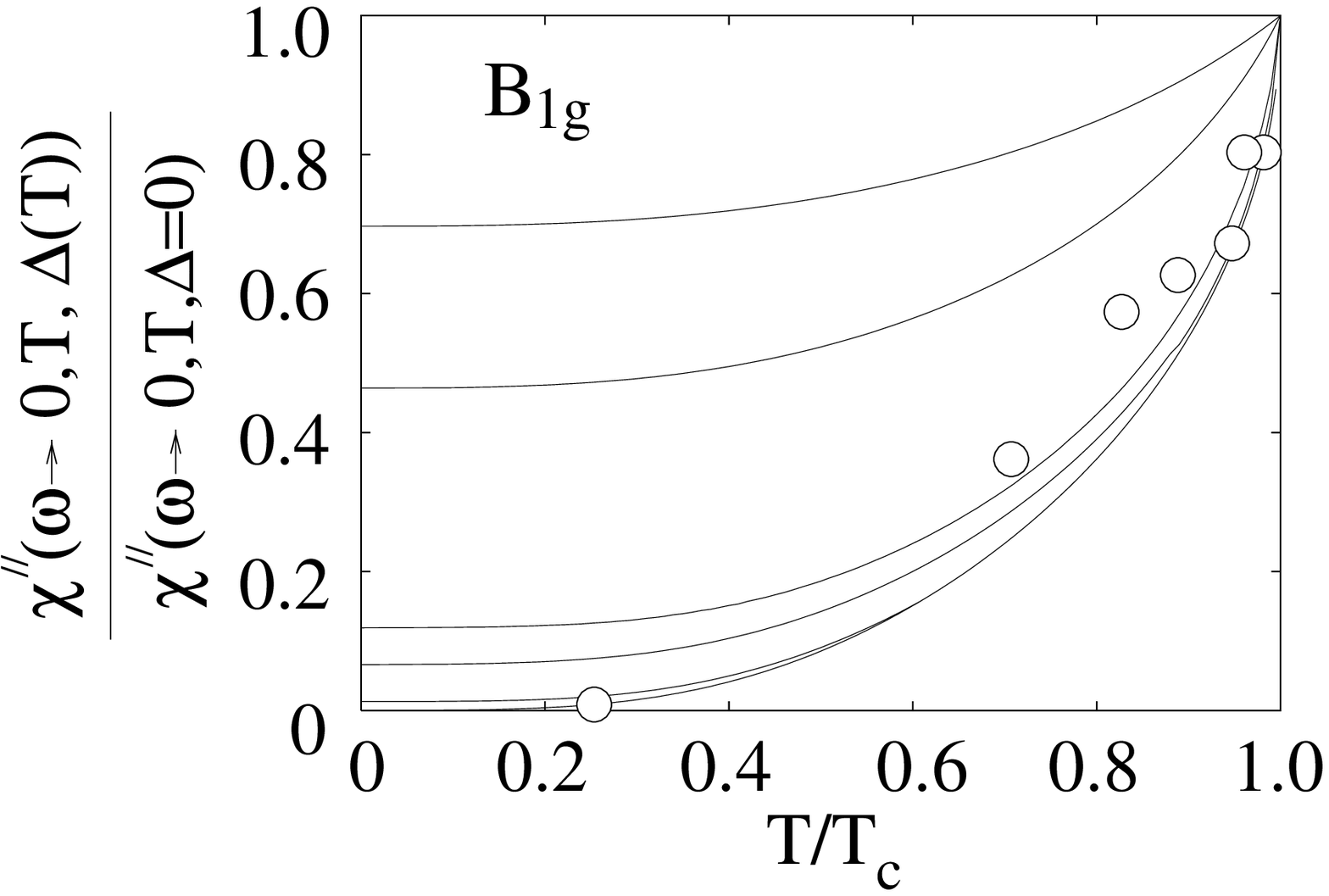,height=6.cm,width=8.cm,angle=0}
\fcaption{$B_{1g}$ static ratio as defined in the text as a function of
progressively stronger disorder ($\Gamma/\Delta(T=0)=0,0.01,0.05,0.1,0.5$ and
1 from bottom to top). The circles are
taken from the data of Hackl {\it et al.} in Ref. 15.}
\label{fig. 7}
\end{figure}

The static ratio for various values of disorder is shown in Figs. 7 and 8 
for unitary scattering for a $d_{x^{2}-y^{2}}$ paired superconductor. 
The results are compared to data taken on optimally doped Bi-2212. The
number this ratio takes for $T=0$ can be calculated analytically. If
we define $\delta$ as $\delta=\Sigma_{0}^{\prime\prime}(\omega \rightarrow
0)$ the ratio becomes
\begin{equation}
{\chi^{\prime\prime}_{s.s.}(\Omega\rightarrow 0, T=0)\over{\chi^{\prime\prime}_{n.s.}(\Omega\rightarrow 0, T=0)}}=\delta^{3}
{\bigl<{\gamma^{2}({\bf k})\over{(\delta^{2}+\Delta^{2}({\bf k}))^{3/2}}}\bigr>\over{\langle\gamma^{2}({\bf k})\rangle}}.
\label{twentythree}
\end{equation}
This relation shows that in the limit of vanishing scattering, $\delta
<< \Delta_{0}$, once again we recover channel dependent exponents. For
the $B_{1g}$ channel, Eq. (\ref{twentythree}) vanishes as 
$(\delta/\Delta_{0})^{3}$
while for $B_{2g}$ it vanishes linearly in $\delta$. 

\begin{figure}[htbp]
\vspace*{13pt}
\psfig{file=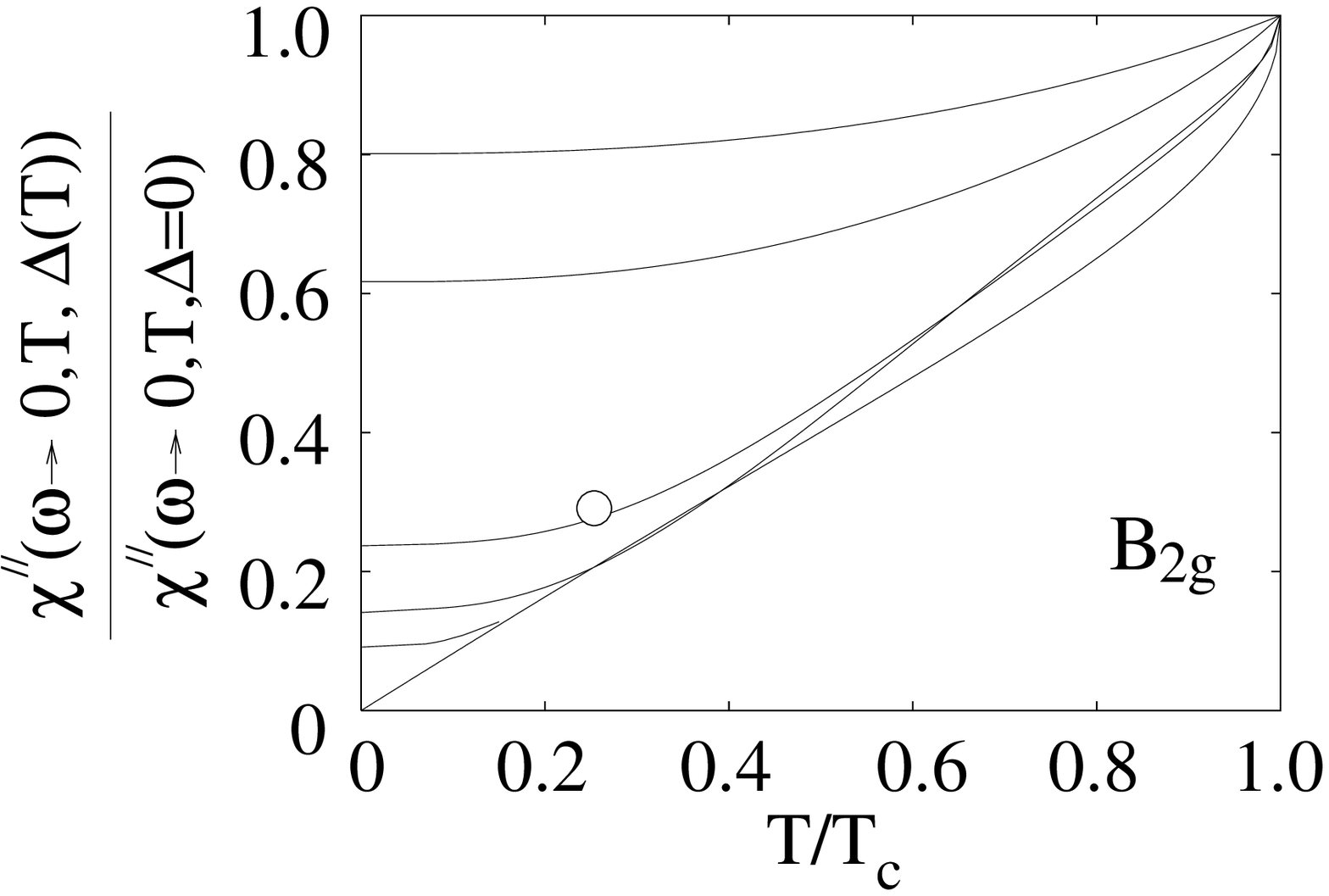,height=6.cm,width=8.cm,angle=0}
\fcaption{$B_{2g}$ static ratio as defined in the text as a function of
progressively stronger disorder ($\Gamma/\Delta(T=0)=0,0.01,0.05,0.1,0.5$ 
and 1 from bottom to top). The circle is
taken from the data of Hackl in Ref. 15.}
\label{fig. 8}
\end{figure}

Finally we determine how large $\delta$ is for both Born and unitary
scattering by solving Dyson's equation self consistently.\cite{pal} For the 
case of Born scattering $c >> 1$, then $\delta \sim e^{\Delta_{0}/\Gamma}$
and is exponentially small. For unitary scattering $c=0$, $\delta$ is
given as a self consistent solution of 
$\delta \sim e^{\delta^{2}}$, which is the Omega function $W$
$\delta=2 \sqrt{\Gamma/\Delta_{0}}/\sqrt{-2W(-\Gamma/\Delta_{0})}$. The
$W$ function provides logarithmic factors and therefore 
$\delta \sim \sqrt{\Gamma/\Delta_{0}}$ in the unitary limit. Therefore
for unitary scattering the constant contribution to the static response
can be quite large and therefore gives a Raman spectrum which produces
normal state behavior with an albeit reduced intensity even at $T=0$.
We note additionally that a constant contribution to the ratio of 
$20\%$ and $33\%$ has been seen for
the $B_{1g}$ and $B_{2g}$ channels, respectively, in Bi-2212 by Yamanaka in
Ref. 5.

In closing this section we remark that all calculations have been performed
using a weak coupling temperature dependence of the energy gap, which does
not provide a good description for the cuprates. However it is trivial to
generalize these results to strong coupling energy gaps by using the
interpolation formula provided in Appendix B of Ref. 9, which
gives the temperature dependence of the gap in analytic form using the
specific heat jump and angular averages of the energy gap as input 
parameters. The above results are only quantitatively changed near $T=T_{c}$.

Moreover, we have performed all calculations for a 2-D cylinder-like
FS and therefore do not capture log corrections that would
occur for all low frequency and temperature behavior of all quantities in
3-D. In the following subsection we will discuss the role of vertex corrections
which have been neglected in this subsection as well.

\subsection{Vertex corrections}

We return to the expression Eq. (\ref{nineteen}) for the
vertex-corrected Raman response and
discuss what changes to the spectra occur when 
impurity vertex corrections are taken into account. In particular we
discuss whether any disorder generated collective modes appear in the
spectra and examine any channel dependence which results from the
vertex corrections. We remind the reader that we are neglecting vertex
corrections from the particle-particle pairing interaction and therefore
these results are not truly speaking gauge invariant even though the
impurities themselves are treated gauge invariantly.

We limit discussion only to the case for $T=0$, where the Matsubara
sum can be converted into an integral and the analytic continuation to
the real frequency axis can be done without having to resort to Pad\'e
approximants or other numerical techniques such as maximum entropy.

We first examine whether there exist any collective modes introduced
by the disorder for a $d_{x^{2}-y^{2}}$ superconductor. We note again 
that the second term in Eq. (\ref{nineteen}) is non-zero only for 
$A_{1g}$ scattering and therefore we only will focus on the third term
which is non-zero only in the $B_{1g}$ channel. Therefore there are no
impurity vertex corrections for the $B_{2g}$ channel and subsequently
no collective modes. 

We have searched for frequencies where the real part of the denominator
is zero in the second term of Eq. (\ref{nineteen}) and have not
been able to identify the existence of any collective modes. We plot
in Fig. 9
the results of the vertex corrected response for the $B_{1g}$ channel
and compare them to the results we obtain neglecting vertex corrections.
We see that the spectra are essentially unmodified apart from a
suppression of spectral weight for frequencies near the gap edge
$2\Delta_{0}$. In particular we do not see a well defined collective mode and
the low and high frequency behaviors remain essentially unchanged. Therefore
the neglect of vertex corrections does not lead to any substantial changes
to the impurity averaged Raman response.
This is similar to the results obtained for the
pair-interaction-vertex-corrected response obtained in Ref. 9, 
where it was shown that the vertex corrections lead to essentially no major
modifications.

\begin{figure}[htbp]
\vspace*{13pt}
\psfig{file=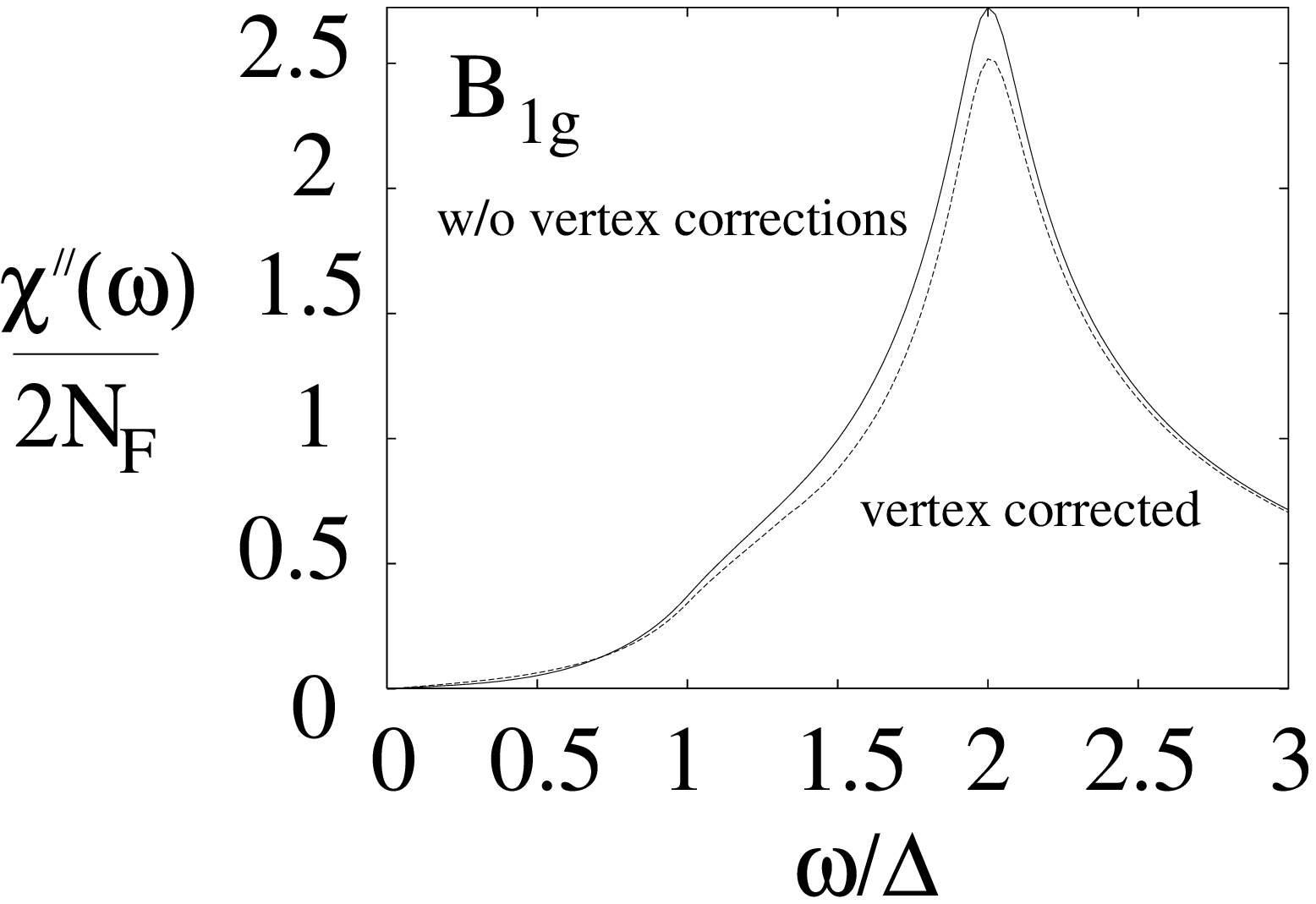,height=6.cm,width=8.cm,angle=0}
\fcaption{Comparison of the calculation of the $B_{1g}$ zero temperature 
spectrum with and without vertex corrections for $\Gamma/\Delta=0.1$ and 
unitary scatterers.}
\label{fig. 9}
\end{figure}

\section{Comparison to data on cuprate superconductors.}

\subsection{Optimally and overdoped cuprates.}

In this section we compile our results and compare them to recent
data taken on the high T$_{c}$ cuprates. We remark at the onset that
no systematic check of the effects of dopant impurities on the electronic
Raman spectra have been undertaken in the metallic state. We do note that
doping of Zn \cite{magnon1} and Pr\cite{magnon2} have been inspected via 
Raman scattering on the two-magnon spectra obtained in insulating and
underdoped Y-123.
There it was seen that the two-magnon feature decreased with the introduction 
of Zn impurities, while a spectral reorganization of the $A_{1g}$ intensity 
was observed at roughly a temperature 20 percent higher than T$_{c}$. Since 
Zn is believed to seriously distort the local antiferromagnetic order in the 
CuO$_{2}$ plane, this finding lends support that antiferromagnetic 
correlations are strong in these compounds for doping levels where 
superconductivity is established. The cause of
the spectral reorganization for Pr doping is unknown at present.

There is recent evidence which suggests that these materials are {\it
intrinsically} disordered in the overdoped regions of their phase diagram. 
Measurements of Raman scattering in the normal state of overdoped Bi-2212
by Hackl {\it et al.} in Ref. 15 
have inferred a larger extrapolated zero temperature scattering rate than
that obtained for optimal doped samples. Moreover the scattering 
rate obtained from the $B_{2g}$ channel matched that obtained from D.C. 
transport.\cite{dctrans} Also, recent muon spin rotation\cite{muon} and Hall
effect data\cite{hall} have suggested that at least for 
overdoped materials
there seems to be 
intrinsic disorder which is manifest in larger quasiparticle
scattering rates. 

One possible scenario of how overdoped materials can be considered to be more
disordered lies in a crossover argument from 2-D to 3-D behavior. In optimally 
doped materials, the carriers responsible for superconductivity are confined to
the 2-D CuO$_{2}$ layers. It is known that these materials become more 
isotropic (c-axis lattice parameter decreases) as they are 
overdoped.\cite{kishio} This is manifest in that the c-axis conductivity 
becomes more and more metallic with overdoping.\cite{homes} As the material 
becomes more 3-D the marginally confined carriers can interact with structural 
distortions in the charge reservoirs which separate the CuO$_{2}$ layers. 
This has been put forth in Ref. 27 as an explanation as to why Ca 
doping Y-123 (which substitutes Y) leads to a much slower decrease in T$_{c}$ 
then Zn doping (which substitutes Cu).

Therefore we will apply the theory as a possible explanation for the 
channel dependent Raman spectra
in overdoped materials. We remark once again that we will
treat the energy gap magnitude as a phenomenological parameter to be used
to fit the position of the $B_{1g}$ Raman peak and assume $d_{x^{2}-y^{2}}$
pairing to be independent of doping\cite{pair}. This is not entirely 
unreasonable since a common feature of the Raman data at {\it any} 
polarization is the
observance of large intensity at frequencies extending towards vanishing
energy transfers, implying the existence of gap nodes. If one were to
assume an isotropic $s-$wave energy gap, an explanation would be required
in order to produce some smearing mechanism which is as large as 30\% of the
energy gap to produce the large scattering at low frequencies. Since
the resolution of the Raman measurements is less than 10 cm$^{-1}$ and
inelastic scattering would be diminished by a factor of $e^{-\Delta_{0}/T}$,
this scenario seems unlikely.

\begin{figure}[htbp]
\vspace*{13pt}
\psfig{file=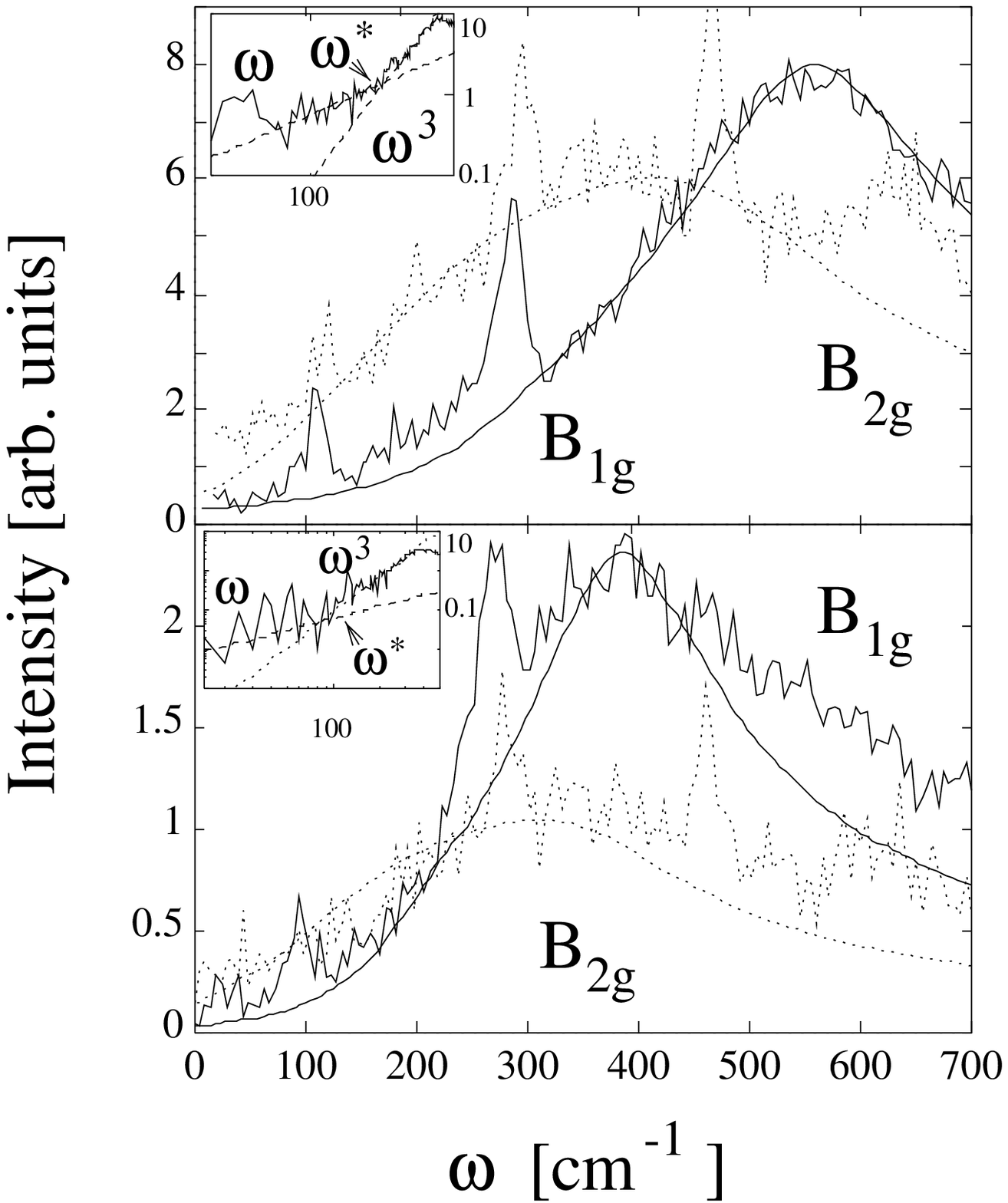,height=10.cm,width=8.cm,angle=0}
\fcaption{Fit to the data taken for $B_{1g}$ and $B_{2g}$ channels in
nearly optimally doped, as-grown Bi 2212 (T$_{c}=86$ K, top panel) and 
slightly overdoped O$_{2}$ annealed Bi 2212
(T$_{c}=79$ K, bottom panel) obtained by Staufer {\it et al.} in Ref.
5. Here $\Gamma/\Delta_{0}= 0.125 (0.2)$ and $\Delta_{0}=
287 (195) \rm{cm}^{-1}$ have been used for the top (bottom) panel,
respectively. Inset: Log-log plot of the low frequency portion of the
$B_{1g}$ response. The crossover frequency $\omega^{*}/\Delta_{0}=
0.38 (0.45)$ for the top (bottom) panels, respectively.}
\label{fig. 10}
\end{figure}

We present a fit of theory in Figs. 10 and 11 to the $B_{1g}$ and $B_{2g}$
spectra obtained on nearly optimally doped (T$_{c}=86 $K), slightly 
overdoped (T$_{c}=79$ K), and appreciably overdoped (T$_{c}=55$ K)
samples of Bi-2212 taken by Staufer {\it et al.} in Ref. 5 and 
by Hackl {\it et al.} in Ref. 15. 
No additional smearing mechanism is invoked. We see that while the
peak positions of the data seem to move together as doping is increased
the low frequency behavior of the combined data is well accounted for
assuming a $d-$wave paired state and an increasing value of the
resonant impurity scattering. This is manifest by the growth of the linear
contribution of the $B_{1g}$- low-frequency response and the persistence of
the $B_{2g}$ linear $\Omega$ behavior. Values of $\Gamma/\Delta_{0}$
estimated by the crossover frequency which separates linear from cubic
behavior in the $B_{1g}$ channel yield $\omega^{*}/\Delta=0.38 (0.45, 0.58)$ 
for the optimally (mildly, appreciably over-) doped samples, respectively. 
Thus spectral weight is being shifted to lower energies as the material
becomes more overdoped. This lends supports to the
conjecture that these materials are more intrinsically disordered for
lower values of T$_{c}$ on the overdoped side of the phase diagram. However
we note that the fit to the appreciably overdoped spectra in Fig. 11, 
especially the $B_{2g}$ spectrum, is not as good as the fits to the other 
spectra. The position of the $B_{2g}$ peak cannot be accounted for 
with our simple theory. Here additional physics seems to
be needed. It may be possible that a
different number of additional harmonics for the gap function (but still
with $d_{x^{2}-y^{2}}$ symmetry) may move the peak outward 
(see Ref. 10 and Branch and Carbotte in Ref. 8). However, without
a detailed theory for the pairing mechanism and its doping dependence this
remains an open question.

\begin{figure}[htbp]
\vspace*{13pt}
\psfig{file=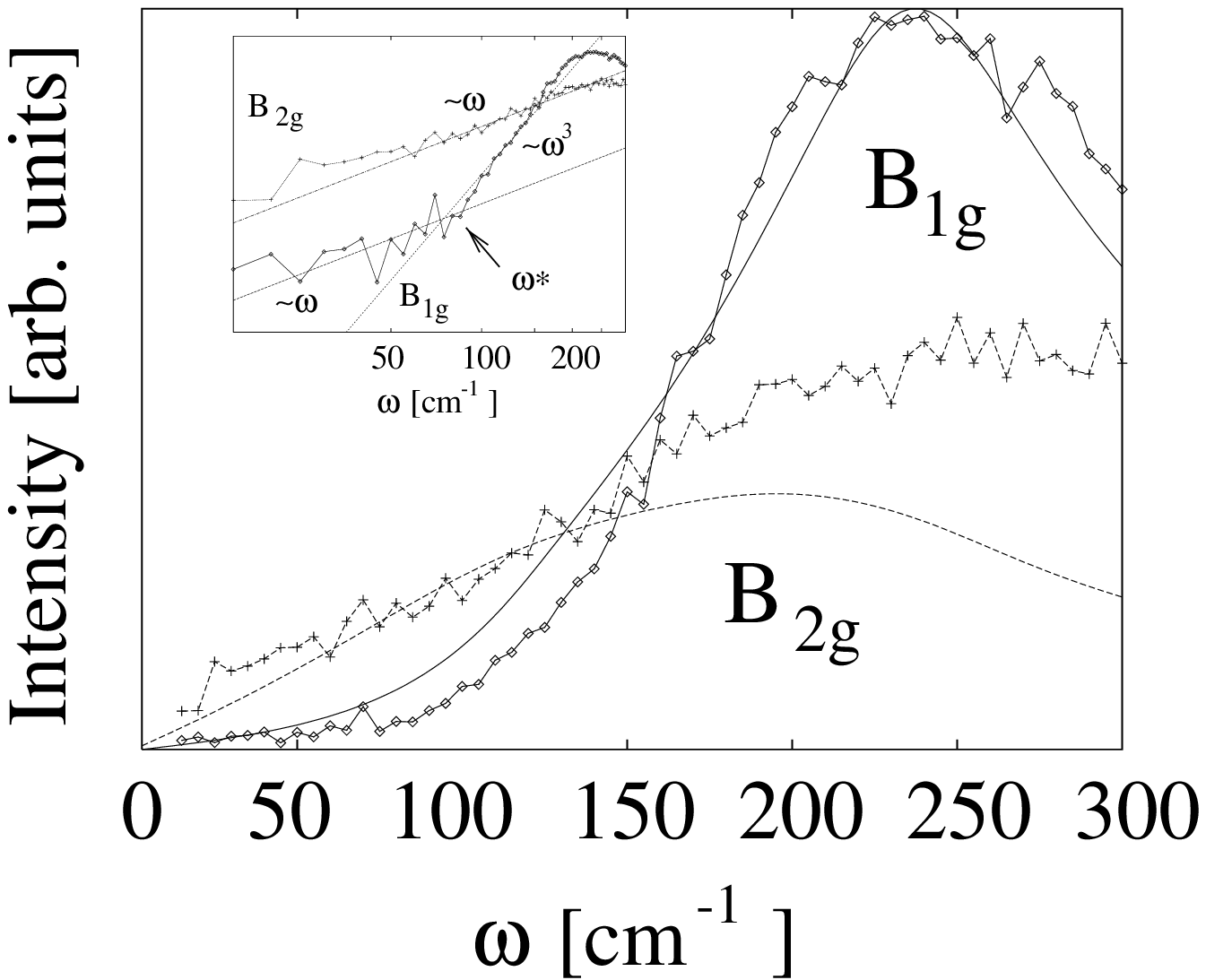,height=8.cm,width=6.cm,angle=0}
\fcaption{Fit to the data taken for $B_{1g}$ and $B_{2g}$ channels in
appreciably doped Bi 2212 (T$_{c}=55$ K) obtained by R. Hackl {\it et al.} 
in Ref. 15. Inset: Log-log plot of the low frequency portion of the
$B_{1g}$ response. Here $\Gamma/\Delta_{0}=0.22$ and $\Delta_{0}=120 
\rm{cm}^{-1}$ have been used. The crossover frequency 
$\omega^{*}/\Delta_{0}=0.58$.}
\label{fig. 11}
\end{figure}

We have neglected quasiparticle inelastic scattering via e.g. spin
fluctuations or phonons which will lead to a further smearing of the
curves and especially the large peak at 2$\Delta_{0}$ in the $B_{1g}$
channel. However, due to the rapid drop off of the scattering rates at
low temperatures and frequencies, the low frequency behavior of the
Raman spectra will not be altered. Inelastic scattering, included 
phenomenologically in the calculations of Ref. 7 as well as
in the calculations of
Jiang and Carbotte in Ref. 8, is needed in order to reproduce
the normal state behavior for temperatures above T$_{c}$ as well as the
flat background seen at much higher energy shifts even at low temperatures.
Thus smaller values of disorder may then be used to fit the data.
Moreover, the effect of resonant Raman scattering will be of more importance
at higher frequencies and may play a pivotal role for a truly microscopic
picture of Raman scattering as a function of doping.

\subsection{Conclusions and open problems.}

We close this section by summarizing our results and listing some open
questions concerning a more complete picture of Raman scattering in the
cuprate materials. We have seen that the theory can provide a satisfactory
fit to the low frequency part of the channel dependent Raman response in
the cuprate superconductors for optimal doping and for overdoping if we
na\"ively assume that the main effect of the doping is to introduce intrinsic
resonant impurity scattering. This is certainly an open question and an
unified picture of what happens even when materials are deliberately
disordered via e.g. Zn or Ni doping is still lacking.\cite{hirsch}
Moreover, the values of scattering needed to fit the Raman data
are indeed quite large to describe the variation of T$_{c}$ as a function of
doping if one applies Abrikosov and Gorkov's theory.\cite{ag} As noted, 
incorporated inelastic scattering could lead to smaller values of disorder
needed to fit the data. However, the
scattering rates implied by the Raman data match those obtained from
D.C. transport, muon spin rotation, and Hall data rather well and do
imply increased scattering on the overdoped side of the phase diagram.
This still leaves the question of why T$_{c}$ is not zero unanswered.

One possible scenario is that the scattering due to impurities is highly
anisotropic. This evidence comes from the data as well, where both the
Raman (see Stadlober {\it et al.} in Ref. 5 and Hackl
{\it et al.} in Ref. 15) and Hall data\cite{hall,stojk} suggest 
that scattering is larger along the BZ faces than along the diagonals.
If this were true, part of the larger scattering could
have the same symmetry as the energy gap and thus not be pair-breaking
in the Anderson sense.\cite{anderson} The theory would have to be changed to
incorporate an extended ${\bf k}-$dependent impurity potential.\cite{tpdak}

This certainly appears to the case on the underdoped side of the phase
diagram which we have not addressed. However, the scattering is most likely
inelastic rather than elastic. Issues of anisotropic inelastic
quasiparticle scattering near ``hot spots'' of the FS have
been invoked to explain the FS evolution\cite{morr} with underdoping 
as well as transport rates obtained from a Boltzmann approach.\cite{stojk} 
It is then
clear that including inelastic quasiparticle scattering is a crucial missing
point of the theory and is needed to explain the Raman spectra on the
underdoped side of the phase diagram. In most Raman experiments, it is
difficult to observe any superconductivity related effects in the $B_{1g}$
channel while they still persists for the $B_{2g}$ response.\cite{doping}
Moreover the relative intensities change as a function of doping in the
normal state. For optimally  and overdoped material, the Raman intensity
in the $B_{1g}$ channel is always larger than the $B_{2g}$ intensity for
any compound measured. However, the $B_{1g}$ intensity has been observed to
drop with underdoping and can become smaller than the $B_{2g}$ 
intensity.\cite{irwin}
This may be related to the loss of the FS around the ``hot spots''
which are most effectively probed in $B_{1g}$ polarization orientations
via the selection rules. Since $B_{2g}$ measures the zone diagonals, the
vestige of the FS, or ``hole pockets'' could still provide
for a large electronic Raman signal. It is thus clear that the role
of electron-electron interactions and incipient antiferromagnetism will
be needed to be incorporated into a theory of Raman scattering for
underdoped materials.

Moreover, recent measurements\cite{magnon1} have shown that the
Raman spectra for all channels in underdoped materials has a resonance
profile that is the same in the normal state as in the superconductor.
Therefore one must include resonant Raman scattering processes into
the theory even for underdoped materials. This is even more important
if one wants to construct a theory for Raman scattering which can be
extended to the insulating state and the two-magnon contribution.

Thus in summary, a theory for Raman scattering in 
unconventional superconductors can be useful to provide insight into the
quasiparticle dynamics in high temperature superconductors for various
regions of their phase diagram. However many issues are left unresolved
and will require further work to build ${\bf k}-$space anisotropies and
electron correlations into the theory as well as resonant scattering
processes. Application of the theory supports recent suggestions that
these materials are intrinsically disordered on the overdoped side of the
phase diagram, and moreover $d_{x^{2}-y^{2}}$ pairing can provide an
adequate description of the data.

\nonumsection{Acknowledgments}
\noindent
T.P.D. would like to thank R. Hackl, G. Krug, M. Opel, R. Nemetschek,
B. Stadlober, J. C. Irwin, K. Hewitt, and J. Naeni for providing their data 
prior to publication and for many useful and insightful discussions. T.P.D.
would also like to thank J. C. Irwin and his colleagues at Simon Fraser 
University, where part of this work was completed. We would also like to 
acknowledge helpful discussions with D. Pines and A. Zawadowski. A.P.K.
acknowledges support through a Heisenberg fellowship of the Deutsche 
Forschungemeinschaft.

\nonumsection{References}


\begin{thebibliography}{000}
\bibitem{pair}
We note that we also are neglecting the important effects of doping and/or 
disorder on the pairing interaction. This undoubtedly plays a large part of the
story concerning T$_{c}$ suppression, but does not affect the details of the
overall lineshape. To proceed further requires a microscopic mechanism for
pairing itself.
\end{thebibliography}
\end{document}